\documentclass[useAMS, usenatbib]{mnras}

\usepackage{graphicx}
\usepackage[all]{hypcap}
\usepackage{amsmath}

\title[Compaction, Depletion and Replenishment on the Star-Forming Main-Sequence]
{The Confinement of Star-Forming Galaxies into a Main Sequence through Episodes of Gas Compaction, Depletion, and Replenishment}

\author[S. Tacchella, A. Dekel, C. M. Carollo, et al.]
{Sandro Tacchella$^{1}$\thanks{E-mail: \href{mailto:sandro.tacchella@phys.ethz.ch}{sandro.tacchella@phys.ethz.ch}}, 
Avishai Dekel$^{2}$, 
C. Marcella Carollo$^{1}$, 
Daniel Ceverino$^{3,4}$, \newauthor
Colin DeGraf$^{2}$,
Sharon Lapiner$^{2}$,
Nir Mandelker$^{2}$,
Joel R. Primack$^{5}$
\\
\\
\\
$^{1}$Department of Physics, Institute for Astronomy, ETH Zurich, CH-8093 Zurich, Switzerland \\
$^{2}$Center for Astrophysics and Planetary Science, Racah Institute of Physics, The Hebrew University, Jerusalem 91904, Israel \\
$^{3}$Centro de Astrobiologia (CSIC-INTA), Ctra de Torrejon a Ajalvir, km 4, 28850 Torrejon de Ardoz, Madrid, Spain \\
$^{4}$Astro-UAM, Universidad Autonoma de Madrid, Unidad Asociada CSIC, E-28049 Madrid, Spain \\
$^{5}$Department of Physics, University of California, Santa Cruz, CA 95064, USA \\
}

\begin{document}

\setlength{\abovedisplayskip}{0.1pt}
\setlength{\belowdisplayskip}{10pt}

\date{\textit{Accepted Version:} \today}

\pagerange{\pageref{firstpage}--\pageref{lastpage}} \pubyear{2016}

\maketitle

\label{firstpage}

\begin{abstract}

Using cosmological simulations, we address the properties of high-redshift star-forming galaxies (SFGs) across their main sequence (MS) in the plane of star-formation rate (SFR) versus stellar mass. We relate them to the evolution of galaxies through phases of gas compaction, depletion, possible replenishment, and eventual quenching. We find that the high-SFR galaxies in the upper envelope of the MS are compact, with high gas fractions and short depletion times (``blue nuggets''), while the lower-SFR galaxies in the lower envelope have lower central gas densities, lower gas fractions and longer depletion times, consistent with observed gradients across the MS. Stellar-structure gradients are negligible. The SFGs oscillate about the MS ridge on timescales $\sim0.4~t_{\mathrm{Hubble}}$ ($\sim1$ Gyr at $z\sim3$). The propagation upwards is due to gas compaction, triggered, e.g., by mergers, counter-rotating streams, and/or violent disc instabilities. The downturn at the upper envelope is due to central gas depletion by peak star formation and outflows while inflow from the shrunken gas disc is suppressed. An upturn at the lower envelope can occur once the extended disc has been replenished by fresh gas and a new compaction can be triggered, namely as long as the replenishment time is shorter than the depletion time. The mechanisms of gas compaction, depletion and replenishment confine the SFGs to the narrow ($\pm0.3$ dex) MS. Full quenching occurs in massive haloes ($M_{\mathrm{vir}}>10^{11.5}~M_\odot$) and/or at low redshifts ($z<3$), where the replenishment time is long compared to the depletion time, explaining the observed bending down of the MS at the massive end.

\end{abstract}


\section{Introduction}\label{sec:Introduction}

Observations of the galaxy population spanning the last 12.5 billion years of cosmic time have revealed a picture in which the majority of star-forming galaxies (SFGs) follow a relatively tight, almost linear relation between star-formation rate (SFR) and stellar mass ($M_{\star}$), also known as the ``main sequence'' (MS) of SFGs \citep[e.g.,][]{brinchmann04, noeske07, noeske07b, daddi07, elbaz07, salim07, whitaker12, speagle14, pannella15}. The SFR increases with $M_{\star}$ as a power law ($\mathrm{SFR}\propto M_{\star}^{\alpha}$ with $\alpha\sim1$) over at least two orders of magnitude ($\sim10^9-10^{11}~M_{\odot}$). Several studies have found that the SFR towards the highest masses ($M_{\star}\ga10^{11}~M_{\odot}$) falls systematically below the value expected for a simple power law relation, effectively lowering the high mass slope of the $\mathrm{SFR}-M_{\star}$ relation towards lower redshifts \citep{rodighiero10, elbaz11, whitaker12, magnelli14, whitaker14, schreiber15}. The most noticeable feature is that the MS relation at any given redshift shows a rather small scatter of $\sigma_{\mathrm{MS}}\sim0.2-0.3~\mathrm{dex}$ \citep{noeske07, whitaker12, speagle14}. 

It is now well established that there is a strong evolution in the normalization of the MS with redshift. The characteristic specific star formation rates ($\mathrm{sSFR}=\mathrm{SFR}/M_{\star}$) of the MS population evolves strongly with redshift, decreasing by a factor of $\sim20$ from $z=2$ to today \citep[e.g.,][]{schreiber15}. All, hydrodynamic simulations of galaxies \citep[e.g., ][]{dave11, dekel13, torrey14, sparre15_MS}, semi-analytical models \citep[e.g., ][]{dutton10, dave12, mitchell14} and analytical models \citep[e.g., ][]{bouche10, dekel13, lilly13_bathtube, forbes14, dekel14_bathtube}, naturally reproduce a correlation between SFR and $M_{\star}$. These studies show that a natural way to understand the decline of the sSFR with time is provided by the predicted decline of gas accretion rate onto the galaxies, which itself is closely related to the evolution of the cosmological specific accretion rate into dark matter haloes, which scales as $\propto(1+z)^{2.5}$ at a fixed mass in the Einstein-deSitter regime, valid at $z>1$ \citep{neistein06, birnboim07, neistein08, fakhouri09, genel10, dutton10, bouche10, tacchella13, lilly13_bathtube, dekel13}.

As SFGs grow in mass, they seem to propagate along the MS, typically not deviating by more than $\pm0.3$ dex from the MS ridge\footnote{The ridge is generally the line connecting the medians of the sSFR at a given stellar mass, or the points of maximum number density of galaxies at the given mass. These two definitions roughly coincide as the distribution about the ridge is roughly symmetric (and log-normal). A detailed definition of the MS ridge is provided by \citet{renzini15}.}, whose sSFR amplitude steadily declines in time. What is the mechanism that keeps the evolving galaxy so tightly confined to the vicinity of the MS ridge until it quenches and falls below the MS? From the cosmological paradigm, dark matter haloes, and hence the central galaxies occupying them, form hierarchically –-- large haloes are built from mergers of smaller haloes. One would therefore expect that mergers between galaxies would frequently trigger starbursts that would generate larger excursions about the MS ridge. However, the small scatter in the MS at multiple redshifts has indicated that most galaxies are not in fact experiencing the expected dramatic effects of major mergers \citep{noeske07, noeske07b, rodighiero11}, and most stars form in ``normal'' galaxies lying along this relation. The SFRs of MS galaxies seem to be sustained for extended periods of time in a quasi-steady state of gas inflow, gas outflow, and gas consumption \citep{daddi10, bouche10, genzel10, tacconi10, dave12, lilly13_bathtube, dekel13, dayal13, feldmann15}. 

Observations indicate that the gas fraction and depletion time tend to vary as a function of sSFR across the MS: high sSFR is correlated with high gas fraction and short depletion time \citep{magdis12, sargent14, huang14, genzel15, silverman15, scoville15}. These gradients may provide a clue for understanding the MS width. Therefore, in this paper, we focus on galaxy properties as a function of sSFR with respect to the MS ridge, rather than the absolute value of sSFR. We define the \textit{universal} MS to be the sSFR with respect to the sSFR of the MS ridge. The main questions that we address in this paper are: (i) What is the mechanism that confines the MS to a small scatter? (ii) What drives the gradients of galaxy properties across the universal MS?

\citet{dutton10} used a semi-analytical model for disc galaxies to explore the origin of the time evolution and scatter of the MS. They find a significant but small scatter in their model MS arising from variation in halo concentration, which in turn causes differences in the mass accretion histories between different galaxies of the same halo mass \citep{wechsler02}. \citet{forbes14} presented a toy model in which the scatter ultimately arises from the intrinsic scatter in the accretion rate, but may be substantially reduced depending on the timescale on which the accretion varies compared to the timescale on which the galaxy loses gas mass. They show that observational constraints on the scatter in galaxy scaling relations can be translated into constraints on the galaxy-to-galaxy variation in the outflow mass loading factor at fixed mass, and the timescales and magnitude of a stochastic component of accretion onto the SFGs. 

The key question is which timescale is encoded in the MS scatter, i.e., does the MS scatter arise because galaxies change their SFR on short timescales ($\sim10^7$ yr), intermediate-timescales ($\sim10^{8-9}$ yr), or long timescales ($\sim10^{10}$ yr) \citep{abramson15, munoz15}. If the MS scatter arises due to short term fluctuations in the star-formation history, similar mass SFGs mostly grow-up together \citep[e.g., ][]{peng10_Cont, behroozi13b}. On the other hand, if the MS scatter arises due to long term fluctuations, similar massive SFGs do not grow-up together and key physics lies in what diversifies star-formation histories \citep[e.g., ][]{gladders13, kelson14}.

How SFGs grow their mass during their life on the MS is also crucial to understand the build up of the quenched population, and the evolution of its median size. \citet{carollo13a} argue indeed for a straight mass-dependent quenching process \citep{peng10_Cont} of M* disc galaxies from the MS to the quenched population (with dry mergers playing a major role in building the $>>$M* quenched population, whose properties point at a dissipationless process as their last step in their assembly histories). A test of this picture is to explore how SFGs grow in mass and size on the MS, and compare the properties of the massive systems that transition to the quenched population at the end of their active lives.

We argue here that the confining mechanism and the gradients across the MS can be understood in terms of the gas regulation, i.e., the balance between inflow rate, SFR and outflow rate, of SFGs at high redshift. We emphasise the importance of internal physical process, likely driven by external events, in addition to global processes (such as gas accretion history). \citet{zolotov15}, analysing cosmological zoom-in simulations, have shown that the processes of gas compaction and subsequent central depletion and quenching are frequent in high-$z$ galaxies and are the major events in their history. They find that stream-fed, highly perturbed, gas-rich discs undergo phases of dissipative contraction into compact, star-forming systems (``blue nuggets''\footnote{In this paper, we refer to compact, SFGs as a blue nuggets. Such galaxies have a high density in their cores, both in stellar mass and gas density. Note that blue nuggets could actually be quiet red due to dust.}) at $z\sim4-2$. The compaction is triggered by an intense inflow episode, involving mergers, counter-rotating streams or recycled gas \citep{dekel14_nugget}, and can be associated with violent disc instability (VDI; \citealt{noguchi99, gammie01, bournaud07, dekel09b, burkert10, bournaud12, cacciato12}). The peak of gas compaction marks the onset of central gas depletion and inside-out quenching.

Here, we try to learn how this characteristic chain of events predicts the gradients across the MS and explains the confinement mechanism. We do this by utilizing the same high-resolution, zoom-in, hydro-cosmological, Adaptive Mesh Refinement (AMR) simulations as \citet{zolotov15}, of galaxies in the redshift range $z=7$ to $z=1$. The suite of 26 galaxies analysed here were simulated at a maximum resolution of $\sim25$ pc including supernova and radiative stellar feedback. At $z\sim2$, the halo masses are in the range $M_{\rm vir}\sim10^{11-12}~M_{\odot}$ and the stellar masses are in the range $M_{\star}\sim10^{9.3-10.8}~M_{\odot}$. With these simulations, we focus on the global physical properties of galaxies on the MS. 

This paper is organized as follows. In Section~\ref{sec:Simulations}, we give a brief overview of the simulations. In Section~\ref{sec:GasContent}, we investigate the gas content of the simulated galaxies. In Section~\ref{sec:MainSequence}, we define the MS, and in Section~\ref{sec:GalaxyProperties}, we determine galaxy properties across the MS. The core of this paper is Section~\ref{sec:Confinment}, where we explain the confinement mechanism of the MS. We discuss implications from our MS paper for the cessation of star formation in galaxies and we highlight several caveats of our analysis in Section~\ref{sec:Discussion}. We summarize our results in Section~\ref{sec:Conclusion}.

%%%%%%%%%%%%%%%%%%%%%%%%%%%%%%%%%
\section{Simulations}\label{sec:Simulations}

\begin{table*}
\centering
\begin{tabular}{@{}lccccccccc}
\multicolumn{10}{c}{{\bf The suite of 26 simulated galaxies.}} \\
\hline
Galaxy & $M_{\rm vir}$ & $M_{\star}$ & $M_{\rm gas}$ & SFR & sSFR & $R_{\rm vir}$ & $R_{\mathrm{M}}$ & $a_{\mathrm{fin}}$ & $z_{\mathrm{fin}}$ \\
  & $10^{12}~M_{\odot} $& $10^{10}~M_{\odot}$ & $10^{10}~M_{\odot}$ & $M_{\odot}/$yr & Gyr$^{-1}$ & kpc & kpc & & \\
  & ($z=2$) & ($z=2$) & ($z=2$) & ($z=2$) & ($z=2$) & ($z=2$) & ($z=2$) & & \\
\hline
\hline
01 & 0.16 & 0.22 & 0.12 & 2.65 & 1.20 & 58.25 & 1.06 & 0.50 & 1.00 \\
02 & 0.13 & 0.19 & 0.16 & 1.84 & 0.94 & 54.50 & 2.19 & 0.50 & 1.00 \\
03 & 0.14 & 0.43 & 0.10 & 3.76 & 0.87 & 55.50 & 1.7 & 0.50 & 1.00 \\
06 & 0.55 & 2.22 & 0.33 & 20.72 & 0.93 & 88.25 & 1.06 & 0.37 & 1.70 \\
07 & 0.90 & 6.37 & 1.42 & 26.75 & 0.42 & 104.25 & 2.78 & 0.50 & 1.00 \\
08 & 0.28 & 0.36 & 0.19 & 5.76 & 1.58 & 70.50 & 0.76 & 0.50 & 1.00 \\
09 & 0.27 & 1.07 & 0.31 & 3.97 & 0.37 & 70.50 & 1.82 & 0.39 & 1.56 \\
10 & 0.13 & 0.64 & 0.11 & 3.27 & 0.51 & 55.25 & 0.53 & 0.50 & 1.00 \\
11 & 0.27 & 1.02 & 0.58 & 17.33 & 1.69 & 69.50 & 2.98 & 0.46 & 1.17 \\
12 & 0.27 & 2.06 & 0.19 & 2.91 & 0.14 & 69.50 & 1.22 & 0.39 & 1.56 \\
13 & 0.31 & 0.96 & 0.98 & 21.23 & 2.21 & 72.50 & 3.21 & 0.39 & 1.56 \\
14 & 0.36 & 1.40 & 0.59 & 27.61 & 1.97 & 76.50 & 0.35 & 0.42 & 1.38 \\
15 & 0.12 & 0.56 & 0.14 & 1.71 & 0.30 & 53.25 & 1.31 & 0.50 & 1.00 \\
20 & 0.53 & 3.92 & 0.48 & 7.27 & 0.19 & 87.50 & 1.81 & 0.44 & 1.27 \\
21 & 0.62 & 4.28 & 0.57 & 9.76 & 0.23 & 92.25 & 1.76 & 0.50 & 1.00 \\
22 & 0.49 & 4.57 & 0.21 & 12.05 & 0.26 & 85.50 & 1.32 & 0.50 & 1.00 \\
23 & 0.15 & 0.84 & 0.19 & 3.32 & 0.39 & 57.00 & 1.38 & 0.50 & 1.00 \\
24 & 0.28 & 0.95 & 0.28 & 4.39 & 0.46 & 70.25 & 1.79 & 0.48 & 1.08 \\
25 & 0.22 & 0.76 & 0.08 & 2.35 & 0.31 & 65.00 & 0.82 & 0.50 & 1.00 \\
26 & 0.36 & 1.63 & 0.25 & 9.76 & 0.60 & 76.75 & 0.76 & 0.50 & 1.00 \\
27 & 0.33 & 0.90 & 0.52 & 8.75 & 0.97 & 75.50 & 2.45 & 0.50 & 1.00 \\
29 & 0.52 & 2.67 & 0.39 & 18.74 & 0.70 & 89.25 & 1.96 & 0.50 & 1.00 \\
30 & 0.31 & 1.71 & 0.41 & 3.84 & 0.22 & 73.25 & 1.56 & 0.34 & 1.94 \\
32 & 0.59 & 2.74 & 0.37 & 15.04 & 0.55 & 90.50 & 2.6 & 0.33 & 2.03 \\
33 & 0.83 & 5.17 & 0.45 & 33.01 & 0.64 & 101.25 & 1.22 & 0.39 & 1.56 \\
34 & 0.52 & 1.73 & 0.42 & 14.79 & 0.85 & 86.50 & 1.9 & 0.35 & 1.86 \\\hline
\end{tabular}
 \caption{Quoted are the total virial mass, $M_{\rm vir}$, the stellar mass, $M_{\star}$, the gas mass, $M_{\rm gas}$, the star formation rate, SFR, the specific star formation rate, sSFR, the virial radius, $R_{\rm vir}$, the effective stellar (half-mass) radius, $R_{\mathrm{M}}$, all at $z=2$, and the final simulation snapshot, $a_{\mathrm{fin}}$, and redshift, $z_{\mathrm{fin}}$. The $M_{\star}$, $M_{g}$, SFR, and sSFR are measured within a radius of $0.2\times R_{\rm vir}$.}
\label{tab:sample}
\end{table*}

We use zoom-in hydro-cosmological simulations of 26 moderately massive galaxies, a subset of the 35-galaxy \texttt{VELA} simulation suite. The details the \texttt{VELA} simulations are presented in \citet{ceverino14_radfeed} and \citet{zolotov15}. \citet{zolotov15} and \citet{tacchella15_profile} used the same sample of 26 simulations and investigated similar questions concerning compaction and quenching. \citet{zolotov15} focused on the evolution of the global properties of the galaxies and their cores as they go through the compaction and quenching phases, and \citet{tacchella15_profile} addresses the evolution of surface density profile of these galaxies during these phases. Additional analysis of the same suite of simulations are discussed in \citet{moody14}, \citet{snyder15} and \citet{ceverino15b}. In this section, we give an overview of the key aspects of the simulations.

\subsection{Cosmological Simulations}

The \texttt{VELA} simulations utilize the Adaptive Refinement Tree (ART) code \citep{kravtsov97, kravtsov03, ceverino09}, which accurately follows the evolution of a gravitating $N$-body system and the Eulerian gas dynamics. All the simulations were evolved to redshifts $z\la2$, and several of them were evolved to redshift $z=1$, with an AMR maximum resolution of $17-35~\mathrm{pc}$ at all times, which is achieved at densities of $\sim10^{-4}-10^3~\mathrm{cm}^{-3}$. In the circumgalactic medium (at the virial radius of the dark-matter halo), the median resolution amounts to $\sim500~\mathrm{pc}$. Beyond gravity and hydrodynamics, the code incorporates the physics of gas and metal cooling, UV-background photoionization, stochastic star formation, gas recycling and metal enrichment, and thermal feedback from supernovae \citep{ceverino10, ceverino12}, plus a new implementation of feedback from radiation pressure \citep{ceverino14_radfeed}. 

We use the CLOUDY code \citep{ferland98} to calculate the cooling and heating rates for a given gas density, temperature, metallicity, and UV background, assuming a slab of thickness 1 kpc. We assume a uniform UV background, following the redshift-dependent \citet{haardt96} model. An exception is at gas densities higher than $0.1~\mathrm{cm}^{-3}$. At these densities, we use a substantially suppressed UV background ($5.9\times10^ 6~\mathrm{erg}~\mathrm{s}^{-1}~\mathrm{cm}^{-2}~\mathrm{Hz}^{-1}$) in order to mimic the partial self-shielding of dense gas, allowing dense gas to cool down to temperatures of $\sim300~\mathrm{K}$. The equation of state is assumed to be that of an ideal mono-atomic gas. Artificial fragmentation on the cell size is prevented by introducing a pressure floor, which ensures that the Jeans scale is resolved by at least 7 cells (see \citealt{ceverino10}).

We assume that star formation occurs at densities above a threshold of $1~\mathrm{cm}^{-3}$ and at temperatures below $10^4~\mathrm{K}$. Most stars ($>90~\%$) form at temperatures well below $10^3~\mathrm{K}$, and more than half of the stars form at $300~\mathrm{K}$ in cells where the gas density is higher than $10~\mathrm{cm}^{-3}$. We use a stochastic star-formation model, where star formation occurs in timesteps of $dt_{\rm SF}=5~\mathrm{Myr}$. The probability to form a stellar particle in a given timestep is

\begin{equation}
P=min\left( 0.2, \sqrt{\frac{\rho_{\rm gas}}{1000~\mathrm{cm}^{-3}}} \right).
\end{equation}
\noindent
The single stellar particle has a mass equal to

\begin{equation}
m_{\star} = m_{\rm gas}\frac{dt_{\rm SF}}{\tau} \approx 0.42m_{\rm gas}
\end{equation}
\noindent
where $m_{\rm gas}$ is the mass of gas in the cell where the particle is being formed and $\tau$ is $12~\mathrm{Myr}$. We assume a \citet{chabrier03} initial mass function. This stochastic star-formation model yields a star-formation efficiency per free-fall time of $\sim2~\%$. At the given resolution, this efficiency roughly mimics the empirical Kennicutt-Schmidt law \citep{kennicutt98}. As a result of the universal local SFR law adopted, the global SFR follows the global gas mass (see Figs. 2 and 3 in \citealt{zolotov15}). Observationally, a universal, local SFR law in which the star formation rate is simply $\sim1~\%$ of the molecular gas mass per local free-fall time fits galactic clouds, nearby galaxies, and high-redshift galaxies \citep{krumholz12a}.

The thermal stellar feedback model releases energy from stellar winds and supernova explosions as a constant heating rate over $40~\mathrm{Myr}$ following star formation. The heating rate due to feedback may or may not overcome the cooling rate, depending on the gas conditions in the star-forming regions \citep{dekel86, ceverino09}. Note that no artificial shutdown of cooling is implemented in these simulations. The effect of runaway stars is included by applying a velocity kick of $\sim10~\mathrm{km}~\mathrm{s}^{-1}$ to $30~\%$ of the newly formed stellar particles. The code also includes the later effects of Type Ia supernova and stellar mass loss, and it follows the metal enrichment of the ISM. 

Radiation pressure is incorporated through the addition of a non-thermal pressure term to the total gas pressure in regions where ionizing photons from massive stars are produced and may be trapped. This ionizing radiation injects momentum around massive stars, pressurizing star-forming regions, as described in Appendix B of \citet{agertz13}. We assume an isotropic radiation field within a given cell and that the radiation pressure is proportional $\Gamma \cdot m_{\star}$, where $m_{\star}$ is the mass of stars and $\Gamma$ is the luminosity of ionizing photons per unit stellar mass. The value of $\Gamma$ is taken from the stellar population synthesis code, \texttt{STARBURST99} \citep{leitherer99}. We use a value of $\Gamma=10^{36}~\mathrm{erg} ~\mathrm{s}^{-1}~M_{\odot}^{-1}$, which corresponds to the time-averaged luminosity per unit mass of the ionizing radiation during the first 5 Myr of evolution of a single stellar population. After 5 Myr, the number high mass stars and ionizing photons declines significantly. Furthermore, the significance of radiation pressure also depends on the optical depth of the gas within a cell. We use a hydrogen column density threshold, $N=10^{21}~\mathrm{cm}^{-2}$, above which ionizing radiation is effectively trapped and radiation pressure is added to the total gas pressure. This value corresponds to the typical column density of cold neutral clouds, which host optically-thick column densities of neutral hydrogen \citep{thompson05}. Summarizing, our current implementation of radiation pressure adds radiation pressure to the total gas pressure in the cells (and their closest neighbours) that contain stellar particles younger than $5~\mathrm{Myr}$ and whose column density exceeds $10^{21}~\mathrm{cm}^{-2}$.

\subsection{Limitation of the current Simulations}\label{subsec:limitations}

The cosmological simulations used in this paper are state-of-the-art in terms of high-resolution AMR hydrodynamics and the treatment of key physical processes at the subgrid level, highlighted above. Specifically, these simulations trace the cosmological streams that feed galaxies at high redshift, including mergers and smooth flows, and they resolve the VDI that governs high-$z$ disc evolution and bulge formation \citep{ceverino10, ceverino12, ceverino15, mandelker14}. 

Like other simulations, the current simulations are not yet doing the best possible job treating the star formation and feedback processes. As mentioned above, the code assumes a SFR efficiency per free fall time that is rather realistic, but it does not yet follow in detail the formation of molecules and the effect of metallicity on SFR \citep{krumholz12b}. Additionally, the resolution does not allow the capture of Sedov-Taylor adiabatic phase of supernova feedback. The radiative stellar feedback assumed no infrared trapping, in the spirit of low trapping advocated by \citet{dekel13a} based on \citet{krumholz12}. Other works assume more significant trapping \citep{murray10, krumholz10, hopkins12}, which makes the assumed strength of the radiative stellar feedback here lower than in other simulations. Finally, AGN feedback and feedback associated with cosmic rays and magnetic fields are not yet incorporated. Nevertheless, as shown in \citet{ceverino14_radfeed}, the star formation rates, gas fractions, and stellar-to-halo mass fractions are all in the ballpark of the estimates deduced from abundance matching, providing a better match to observations than earlier simulations. 

The uncertainties and any possible remaining mismatches between simulation and observations by a factor of order 2 are comparable to the observational uncertainties. For example, the stellar-to-halo mass fraction is not well constrained observationally at $z\sim2$. Recent estimates by \citet{burkert15} (see their Fig. 5) based on the observed kinematics of $z\sim0.6 - 2.8$ SFGs reveal significantly larger ratios than the estimates based on abundance matching \citep{conroy09, moster10, moster13, behroozi10, behroozi13b} at $M_{\rm vir}<10^{12}~M_{\odot}$. In Section~\ref{App:stellarhalo} of the appendix, we present a detailed comparison of the stellar-to-halo mass relation of our simulations and the observational data (Figure~\ref{FigApp:stellarhalo}). We conclude that our simulations produce stellar-to-halo mass ratios that are in the ballpark of the values estimated from observations, and within the observational uncertainties, thus, indicating that our feedback model is adequate.

It seems that in the current simulations, the compaction and the subsequent onset of quenching occur at cosmological times that are consistent with observations (see Fig. 12 of \citealt{zolotov15} and Fig. 2 of \citealt{barro13}). With some of the feedback mechanisms not yet incorporated (e.g., AGN feedback), full quenching to very low sSFR values may not be fully reached in many galaxies by the end of the simulations at $z\sim1$. In this work, we adopt the hypothesis that the simulations grasp the qualitative features of the main physical processes that govern galaxy evolution.

\subsection{Galaxy Sample and Properties}\label{subsec:sample}

The initial conditions for the simulations are based on dark-matter haloes that were drawn from dissipationless N-body simulations at lower resolution in three comoving cosmological boxes (box-sizes of 10, 20, and 40 Mpc/h). The assumed cosmology is the standard $\Lambda$CDM model with the WMAP5 values of the cosmological parameters, namely $\Omega_m=0.27$, $\Omega_{\Lambda}=0.73$, $\Omega_b=0.045$, $h=0.7$ and $\sigma_8=0.82$ \citep{komatsu09}. Each halo was selected to have a given virial mass at $z = 1$ and no ongoing major merger at $z=1$. This latter criterion eliminates less than $10~\%$ of the haloes, which tend to be in a dense proto-cluster environment at $z\sim1$. The target virial masses at $z=1$ were selected in the range $M_{\rm vir}=2\times10^{11}-2\times10^{12}~M_{\odot}$, about a median of $4.6\times10^{11}~M_{\odot}$. If left in isolation, the median mass at $z=0$ was intended to be $\sim10^{12}~M_{\odot}$. In practice, the actual mass range is broader, with some of the haloes merging into more massive haloes that eventually host groups at $z=0$.

From the suite of 35 galaxies, we have excluded three low-mass galaxies that have a quenching attempt that brings them considerably below the MS ($>0.8$ dex) for a short period before they recover back to the MS. This is probably a feature limited to very low mass galaxies, and we do not address it any further here. Furthermore, we have excluded six galaxies which have not been simulated down to $z=2.0$. Therefore, our final sample consists of 26 galaxies. The virial and stellar properties are listed in Table~\ref{tab:sample}. This includes the total virial mass $M_{\rm vir}$, the galaxy stellar mass $M_{\star}$, the gas mass $M_{\rm gas}$, the SFR, the sSFR, the virial radius $R_{\rm vir}$, and the effective, and half-mass radius $R_{\mathrm{M}}$, all at $z=2$. The latest time of analysis for each galaxy in terms of the expansion factor, $a_{\rm fin}$, or redshift, $z_{\rm fin}$, is provided.

The virial mass $M_{\rm vir}$ is the total mass within a sphere of radius $R_{\rm vir}$ that encompasses an overdensity of $\Delta(z)=(18\pi^2-82\Omega_{\Lambda}(z)-39\Omega_{\Lambda}(z)^2)/\Omega_{m}(z)$, where $\Omega_{\Lambda}(z)$ and $\Omega_{m}(z)$ are the cosmological parameters at $z$ \citep{bryan98, dekel06}. The stellar mass of the galaxy, $M_{\star}$, is the instantaneous mass in stars (after the appropriate mass loss), measured within a sphere of radius $0.2\times R_{\rm vir}$ about the galaxy centre. The gas mass, $M_{\rm gas}$, is the cold gas mass within the same sphere, i.e., the gas with a temperature below $10^4$ K. Measuring these global quantities within different radii ($0.1-0.3\times R_{\rm vir}$ or even fixing it to 10 kpc) does not change the main findings of this paper. Throughout this paper, all the quoted gas properties refer to the cold gas component. The effective radius $R_{\mathrm{M}}$ is the three-dimensional half-mass radius corresponding to $M_{\star}$.

The SFR is obtained by $\mathrm{SFR}=\langle M_{\star}(t_{\rm age}<t_{\rm max})/t_{\rm max} \rangle_{t_{\rm max}}$, where $M_{\star}(t_{\rm age}<t_{\rm max})$ is the mass in stars younger than $t_{\rm max}$ within a sphere of radius $0.2\times R_{\rm vir}$ about the galaxy centre. The average $\langle\cdot\rangle_{t_{max}}$ is obtained for $t_{\rm max}$ in the interval $[40,80]~\mathrm{Myr}$ in steps of 0.2 Myr in order to reduce fluctuations due to a $\sim5$ Myr discreteness in stellar birth times in the simulation. The $t_{\rm max}$ in this range are long enough to ensure good statistics. 

We start the analysis at the cosmological time corresponding to expansion factor $a=0.125$ (redshift $z=7$). At earlier times, the fixed resolution scale typically corresponds to a larger fraction of the galaxy size, so the detailed galaxy properties may be less accurate. As visible in Table~\ref{tab:sample}, most galaxies reach $a=0.50$ ($z=1$). Each galaxy is analysed at output times separated by a constant interval in $a$, $\Delta a=0.01$, corresponding at $z=2$ to $\sim100~\mathrm{Myr}$ (roughly half the orbital time at the disc edge). For six galaxies, namely 11, 12, 14, 25, 26, and 27, $\sim20$-times higher resolution snapshots are available, for which the output times separated by $\Delta a=0.0005-0.0007$. In Appendix~\ref{App:thin}, we show that our standard snapshots are tracing the main evolutionary pattern, and the high temporal resolution snapshots confirm our main findings.

%%%%%%%%%%%%%%%%%%%%%%%%%%%%%%%%%
\section{Gas Content in Simulated Galaxies}\label{sec:GasContent}

\begin{figure}
\includegraphics[width=\linewidth]{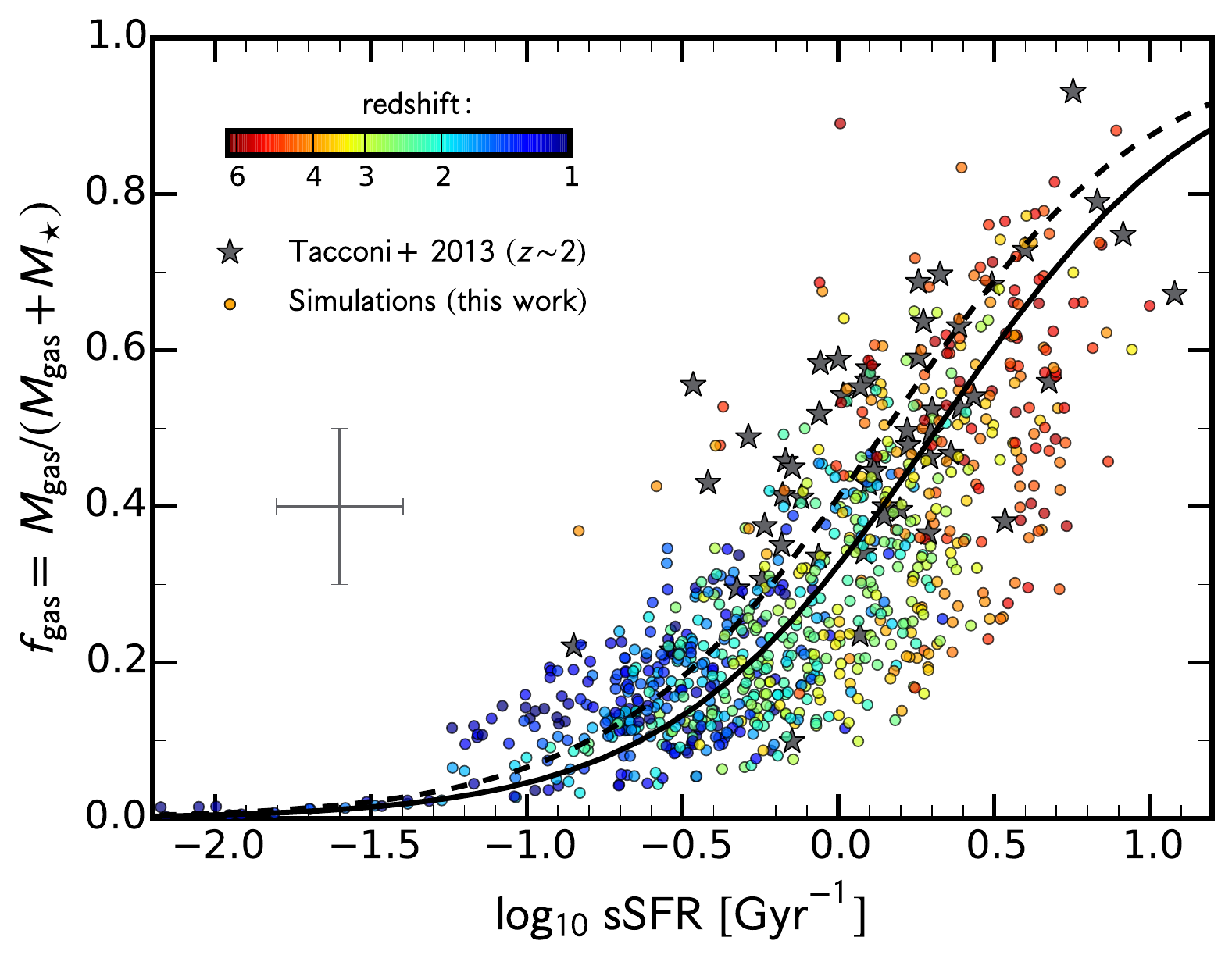} 
\caption{Gas fraction versus sSFR. The circles correspond to all the snapshots of all the galaxies in the simulated sample. The colour coding corresponds to redshift according to the colour bar. We see that $f_{\rm gas}$ and sSFR are systematically decreasing with cosmic time. The simulations are compared to the observations of \citet{tacconi13}, marked by stars. The errorbar indicates the uncertainty in the observations. The simulated and observed galaxies span a similar locus in this plane, despite the fact that the simulations somewhat underestimate both the gas fraction and the sSFR (see text). The black solid line indicates the best-fitting depletion time at $z\leq2.5$ of the simulations ($t_{\mathrm{dep}}=480\pm10~\mathrm{Myr}$), which is shorter than the observational estimate of $t_{\mathrm{dep}}=700\pm200~\mathrm{Myr}$ shown as a black dashed line.}
\label{Fig:GasMass}
\end{figure}

As mentioned before, the star formation in the simulations is driven by the content of dense and cold gas in the galaxies. It is therefore important to investigate the gas properties of our simulated galaxies, before analysing the evolution and shape of the MS. Figure~\ref{Fig:GasMass} shows the gas fraction, $f_{\rm gas}=M_{\rm gas}/(M_{\rm gas}+M_{\star})$, as a function of sSFR. The individual points refer to the individual snapshots, ranging from $z=6$ to 1 as indicated by the colour coding. The high redshift ($z>3$) galaxies of typical masses below $\sim10^{9.5}~M_{\odot}$ and $\mathrm{sSFR}\approx2~\mathrm{Gyr}^{-1}$ tend to have a high gas fraction, $f_{\rm gas}>0.4$. Towards lower redshifts ($z\sim1-3$), when the galaxies in our sample also grow to higher masses, the $f_{\rm gas}$ and the sSFR continuously decline with increasing cosmic time. At $z\sim2$, our simulated $10^{10}~M_{\odot}$ galaxies have $f_{\rm gas}=0.20^{+0.14}_{-0.09}$ (see below). 

The quantities $f_{\rm gas}$ and sSFR are related through

\begin{equation}
f_{\rm gas}=\frac{M_{\rm gas}}{M_{\rm gas}+M_{\star}}=\frac{1}{1+(t_{\mathrm{dep}}\times\mathrm{sSFR})^{-1}},
\label{eq:tdep_def}
\end{equation} 
\noindent
where $t_{\mathrm{dep}}=M_{\rm gas}/\mathrm{SFR}$ is the depletion time\footnote{The depletion time (sometimes also called the gas consumption time) can be written as $t_{\mathrm{dep}} = M_{\rm gas}/\mathrm{SFR} = t_{\mathrm{ff}}/\varepsilon_{\mathrm{ff}}$, where $\varepsilon_{\mathrm{ff}}$ is the SFR efficiency and $t_{\rm ff}$ is the free-fall time in the star-forming regions. The SFR efficiency can be argued to be constant in all star-forming environments, $\varepsilon_{\mathrm{ff}}\sim0.01$, such that the variation in $t_{\mathrm{dep}}$ mostly reflects variations in $t_{\mathrm{ff}}$ \citep{krumholz12a}.}. From Figure~\ref{Fig:GasMass}, we can estimate a global $t_{\mathrm{dep}}$ by fitting all our simulated galaxies at all redshifts together. We find $t_{\mathrm{dep}}=400\pm10~\mathrm{Myr}$ in the simulations. Considering only galaxies at $z\leq2.5$, we find $t_{\mathrm{dep}}=480\pm10~\mathrm{Myr}$. As a by-product to our current analysis, from the fact that this timescale is much shorter than the Hubble time for all SFGs between $z=1-4$, we can conclude that fresh gas must be supplied with a fairly high duty cycle over several billion years.

We compare the simulations with measurements of \citet{tacconi13} who present a CO $3-2$ survey of molecular gas properties in massive galaxies at $z\sim1-3$. They provided 52 CO detections in galaxies with $\log_{10}~M_{\star}/M_{\odot}>10.4$ and $\log_{10}~\mathrm{SFR}/(M_{\odot}~\mathrm{yr}^{-1})>1.5$. The galaxies were selected to sample the complete range of star formation rates within the aforementioned stellar mass limit, ensuring a relatively uniform sampling of the MS. They infer average gas fractions of $f_{\rm gas}\sim0.33$ at $z\sim1.2$ and $\sim0.47$ at $z\sim2.2$ for the given masses, and an overall drop in $f_{\rm gas}$ with $M_{\star}$ at a given redshift. 

Both the observed and the simulated galaxy sample are incomplete and the comparison between the two has to be interpreted with caution. Comparing the \citet{tacconi13} measurements with our simulations, we find that galaxies in the simulations have an average gas deficit at a given stellar mass of $20\%-40\%$ in comparison with the observations. However, at a given sSFR, we find good agreement between observations and simulations (Figure~\ref{Fig:GasMass}). \citet{tacconi13} inferred an average depletion time $t_{\mathrm{dep}}=700\pm200~\mathrm{Myr}$ at $z=1-3$, which may be slightly larger than our estimate of $\sim480$ Myr from the simulations. The good agreement is found because at a given $M_{\star}$, the SFR in the simulations is also underestimated by a similar multiplicative factor as the gas density. We conclude that although the simulated galaxies may evolve a bit ahead of cosmic time, they nevertheless reproduce the qualitative trends seen in the observations.

%%%%%%%%%%%%%%%%%%%%%%%%%%%%%%%%%

\begin{figure*}
\includegraphics[width=\textwidth]{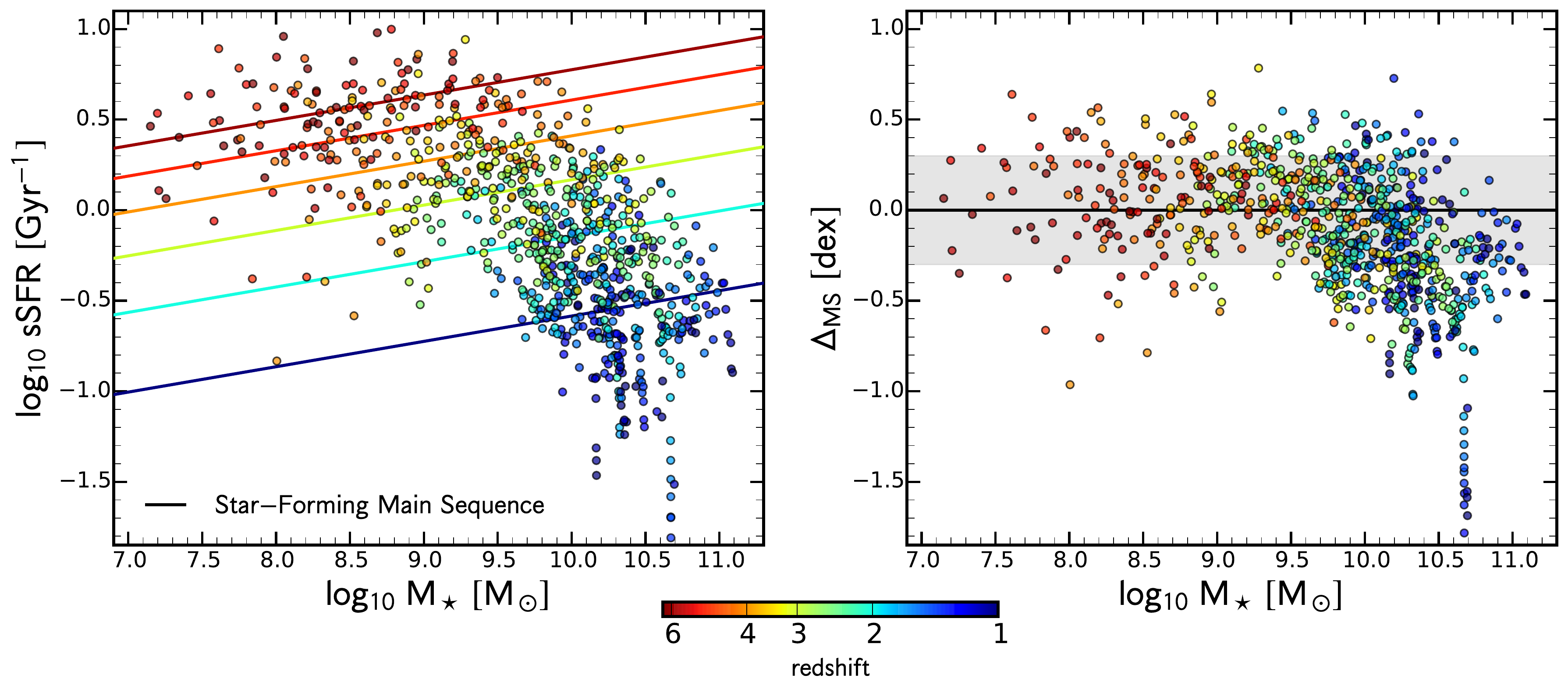} 
\caption{Evolution of the MS in the simulations and the universal MS. The colour refers to redshift according to the colour bar. \textit{Left panel:} sSFR as a function of stellar mass $M_{\star}$. The individual points show the galaxies from the simulations at different redshifts ($z=6-1$). The solid lines show the evolution of the MS ridge sSFR$_{\mathrm{MS}}$ (Equation~\ref{eq:MS}). \textit{Right panel:} The universal MS, i.e., the distance from the MS ridge, $\Delta_{\mathrm{MS}}\equiv\log_{10}(\mathrm{sSFR}/\mathrm{sSFR}_{\mathrm{MS}})$, as a function of $M_{\star}$ and $z$. The solid line marks the MS ridge and the gray shaded region indicates the $\pm0.3$ dex scatter of the MS. The simulations recover the confinement to a narrow MS, with a bending down at the massive end at late times. } 
\label{Fig:Main_Sequence}
\end{figure*}

\section{The Star-Forming Main-Sequence}\label{sec:MainSequence}

In this section, we identify the MS in the simulations and determine the strong systematic time evolution of its ridge and its weak mass dependence, sSFR$_{\mathrm{MS}}$($M_{\star},z$). We are motivated by the hypothesis that the sSFR$_{\mathrm{MS}}$($M_{\star},z$) roughly follows the average specific accretion rate of mass into dark-matter haloes, and test this hypothesis. After obtaining the best-fitting MS in the simulations, we compare it with observations. 

\subsection{Definition of the MS in the Simulations}

The average specific accretion rate of mass into haloes of mass $M_{\rm h}$ at $z$ can be approximated by an expression of the form:

\begin{equation}
\frac{\dot{M_{\rm h}}}{M_{\rm h}} \simeq s_{\rm h} \cdot M_{12}^{\beta} \cdot \left(1+z\right)^{\mu},
\label{eq:sAR}
\end{equation}
\noindent
where $M_{12}=M_{\rm h}/10^{12}M_{\odot}$. In the $\Lambda$CDM cosmology in the Einstein-deSitter regime (valid at $z>1$), simple theoretical arguments show that $\mu\to5/2$ (see \citealt{dekel13}). \citet{neistein08} showed that, for the $\Lambda$CDM power-spectrum slope on galactic scales, $\beta$ is small, $\beta \simeq 0.14$. An appropriate value for the normalization factor $s_{\rm h}$ provides a match better than 5\% at $z>1$ to the average evolution in cosmological N-body simulations \citep{dekel13}. 

To constrain the evolution of the MS ridge with cosmic time, we adopt for the sSFR the same functional form as in Equation~\ref{eq:sAR},

\begin{equation}
\mathrm{sSFR}_{\mathrm{MS}}(M_{\star}, z)=s_{\rm b}\cdot\left(\frac{M_{\star}}{10^{10}~M_{\odot}}\right)^{\beta}\cdot \left(1+z\right)^{\mu}~\mathrm{Gyr^{-1}}.
\label{eq:MS}
\end{equation}
\noindent
The three free parameters ($s_{\rm b}$, $\beta$, and $\mu$) are not independent of each other. In particular, the time evolution of galaxies is characterized by both $\beta$ (stellar mass dependence) and $\mu$ (redshift evolution). We therefore assume $\beta=0.14$ and $\mu=5/2$, i.e., the values of the specific halo mass accretion rate. This choice of $\beta$ and $\mu$ is motivated by our data. Fitting $\beta$ in bins of redshift (bins of 0.5 in the range $z=6-3$), we find as best-fitting $\beta=0.12\pm0.06$. Using $\beta=0.12$ and fitting $\mu$, we find $\mu=2.5\pm0.6$. 

Since we adopt $\beta=0.14$ and $\mu=5/2$, the only free parameter in the MS-scaling is the overall normalisation which we parametrise with $s_{\rm b}$. We determine $s_{\rm b}$ by a least-square fit to $\log_{10}\mathrm{sSFR}$ using all the snapshots in the redshift range $6-3$ for all the galaxies in our sample, excluding all galaxies that are below the 16\%-percentile or above the 84\%-percentile at a given redshift, thus focusing on the inner $\pm1\sigma$ about the median. The choice of this redshift range is motivated by the fact that all of our galaxies in the range $z=6-3$ are star-forming. We find $s_{\rm b}=0.046\pm0.002$. Alternatively, if we replace the redshift cut by a cut in stellar surface density within the inner 1 kpc, $\Sigma_{M_{\star},1\mathrm{kpc}}<10^9~M_{\odot}$, as in \citet{zolotov15}, we obtain a value of $s_{\rm b}$ that is consistent within the uncertainty with the above value, which we adopt here.

Figure~\ref{Fig:Main_Sequence} shows the evolution of the best-fitting MS ridge in the simulations. The left panel shows the MS ridge with solid lines at redshifts $z=[1.0, 2.0, 3.0, 4.0, 5.0, 6.0]$, indicated by the different colours following the colour bar. The individual points refer to all the snapshots of all the sample galaxies at redshifts $1\leq z\leq6$. The right panel of Figure~\ref{Fig:Main_Sequence} shows the universal MS, namely all snapshots after scaling the sSFR according to the scaling of the MS ridge. The figure thus shows the distance of each galaxy from the MS ridge, $\Delta_{\mathrm{MS}}=\log_{10}\left(\mathrm{sSFR}/\mathrm{sSFR}_{\mathrm{MS}}\right)$. The grey shaded area indicates a scatter of $\pm0.3$ dex, indicative of the scatter of the MS. 

Two low-mass galaxies become sub-MS at $z\sim4-3$ and then return to the MS (galaxies 10 and 24). These galaxies live through a short ($<200~\mathrm{Myr}$) phase of nearly no accretion of gas that reduces their SFR significantly. This quenching attempt is terminated by a sudden accretion of fresh gas, usually initiated by a merger. This brings the galaxy back into the MS within less than 400 Myr. Such episodes of quenching attempts seem to occur more frequently in low-mass galaxies at high-$z$, but they are rare above the mass threshold adopted for the sample analysed in this paper (Section~\ref{subsec:sample}). 

At later epochs ($z\la2$), for several galaxies, the quenching process is continuous over several 100 Myr up to several Gyr, indicating a successful quenching as opposed to a short-term quenching attempt at high-$z$. Only 3 galaxies fall $1~\mathrm{dex}$ below the MS by $z=1$, i.e. quench their star-formation substantially. When these galaxies fall below the MS, their stellar mass roughly stays constant. As mentioned in Section~\ref{subsec:limitations}, full quenching to very low sSFR is not reached because the feedback may still be underestimated (e.g., the adiabatic phase of supernova feedback is not resolved, and AGN feedback is not yet incorporated). However, 12 out of our 26 galaxies drop to more than 1 $\sigma$ below the MS ridge by $z=1$, and all are continuously moving downward in $\Delta_{\mathrm{MS}}$ for the last several Gyr, indicating that they are in a long-term quenching process.

\subsection{Comparison of the MS in the Simulations with Observations}

\begin{figure}
\includegraphics[width=\linewidth]{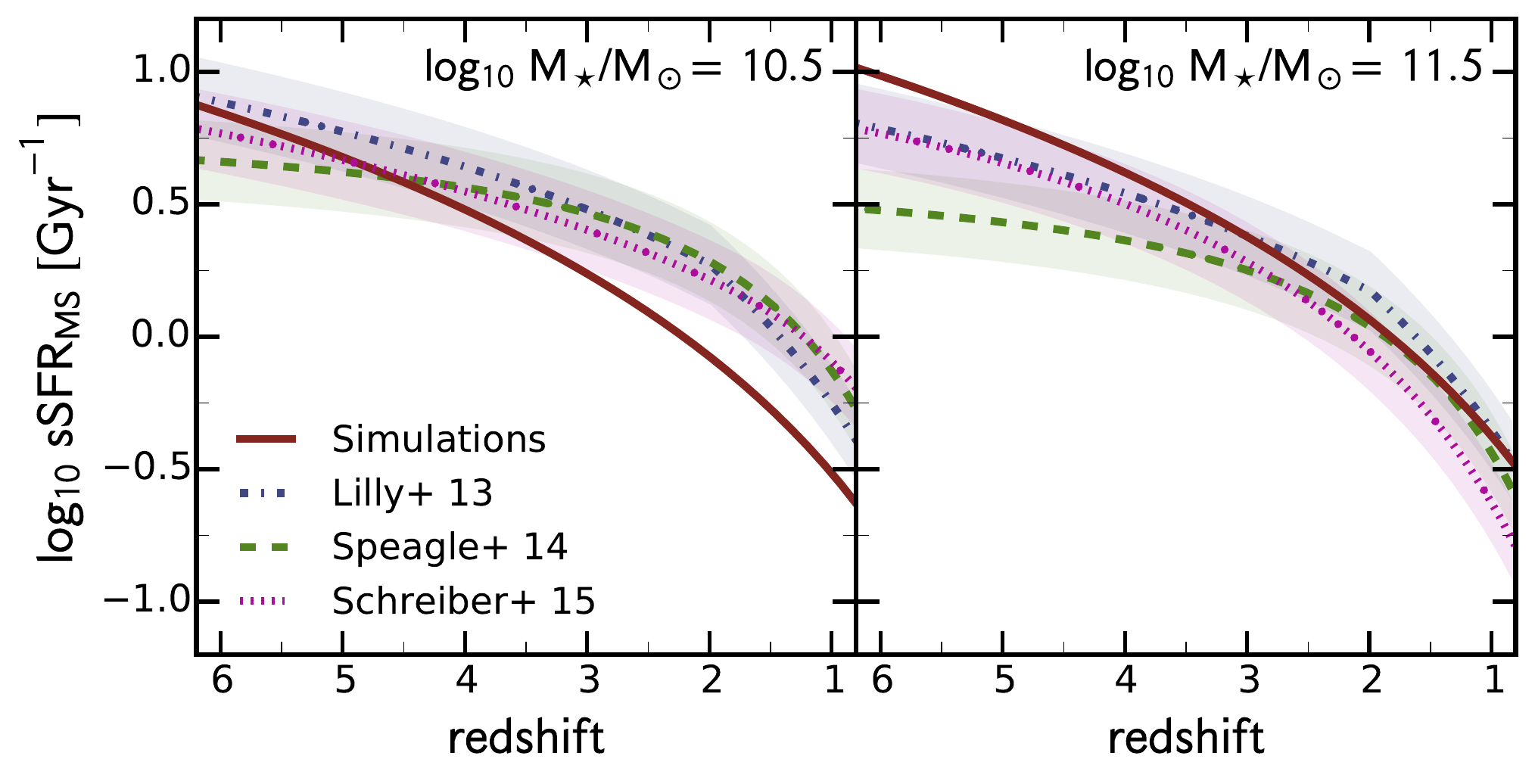} 
\caption{Comparison of the sSFR in the MS ridge from the simulations with observations for galaxies at two mass bins ($\log_{10}M_{\star}/M_{\odot}\simeq10.5$ in the left and extrapolated to 11.5 in the right panel). The red solid line shows the MS ridge from our simulations, i.e., the best-fitting $\mathrm{sSFR_{\mathrm{MS}}}$ (Equation~\ref{eq:MS}). The blue dashed-dotted, green dashed, and purple dotted lines show the best-fitting MS ridges of the observations by \citet{lilly13_bathtube}, \citet{speagle14}, and \citet{schreiber15}, respectively. The shade regions enclosing the lines of the observations with a 0.3 dex width show the approximate observational uncertainties. At $z \sim 2$, the simulations match the observations in the high-mass end, but they are an underestimate by 0.3 dex for the lower masses. }
\label{Fig:Difference_MS}
\end{figure}

Figure~\ref{Fig:Difference_MS} compares the sSFR amplitude of the MS ridge from the simulations with the one deduced from observations \citep{lilly13_bathtube, speagle14, schreiber15} for two different stellar masses. Our simulations, the red curves, show the best-fitting MS introduced above (Equation~\ref{eq:MS}), extrapolated to masses and redshifts beyond what we actually have in the simulations (e.g., we do not have any $10^{11}~M_{\odot}$ galaxies at $z>2$ in the simulations). The \citet{lilly13_bathtube} line is based on best-fitting to the cumulated data of \citet{noeske07, elbaz07, daddi07, pannella09} and \citet{stark13}. The line of \citet{speagle14} uses a compilation of 25 observational studies from the literature out to $z\sim6$. After converting all observations to a common set of calibrations, they find a remarkable consensus among MS ridge observations, with $\sim0.1~\mathrm{dex}$ for the $1~\sigma$ inter-publication scatter. \citet{schreiber15} presented an analysis of the deepest Herschel images obtained within the GOODS-Herschel and CANDELS-Herschel key programs. This allowed them to measure SFR based on direct ultraviolet and far-infrared ultraviolet-reprocessed light and to determine the evolution of the MS at $z=0-4$.

In Figure~\ref{Fig:Difference_MS}, at high masses (right panel), the MS from observations and our simulations are consistent, especially at intermediate redshifts, $z \sim 1-3$. The differences at higher redshifts may reach the levels of $\sim0.3~\mathrm{dex}$, i.e., of the same level as potential systemic errors in both SFR and $M_{\star}$ deduced from observations, and comparable to the difference between the different compilations of observations. At lower masses (left panel), the MS of the simulations is in less good agreement with observations, especially at $z=1-3$, where the simulations lie $0.2-0.4$ dex below the observational estimates.

The larger difference towards lower stellar masses between the MS of the simulations and the observations can be partly explained by a difference in the logarithmic slope of the mass dependence: $\beta=0.14$ for our simulations (and the dark matter haloes, e.g., \citealt{neistein08}), where $\beta_{\rm obs}\sim-0.1$ to 0.0 in the observations \citep[e.g.,][]{schreiber15}. This small difference may be relevant for some physical processes \citep[e.g.,][]{bouche10}, but for our purpose, given the rather limited mass range in our simulated sample, it does not make a difference for the nature of the evolution about the MS. Despite these differences between observed and simulated MS, we hypothesize that a qualitative study of the physical processes governing the evolution of galaxies with respect to the MS ridge can be meaningful once treated self-consistently using the MS as defined in the simulations above.

As noted before, the cosmological rates of evolution of the average observed and simulated sSFR are consistent with the average specific accretion rate of mass into haloes as expressed in Equation~\ref{eq:sAR} \citep{bouche10, dekel13, lilly13_bathtube}. However, there are certain differences between these two quantities. The sSFR (with $s_{\rm b}\approx0.046$ Gyr$^{-1}$) appears to be somewhat higher than the specific halo mass accretion rate (sAR, with $s_{\rm h}\approx0.03$ Gyr$^{-1}$, as measured from another set of simulations in \citealt{dekel13}) over a wide range of redshifts, indicating a factor of $\sim1.5$ shorter $e$-folding time for the build up of stars compared with that of the dark matter haloes. This is consistent with predictions from a bathtub toy model by \citet{lilly13_bathtube} and \citet{dekel14_bathtube} that predict $\mathrm{sSFR}\sim1/f_{\rm star}\cdot\mathrm{sAR}\sim1.5\cdot\mathrm{sAR}$ at high $z$, where $f_{\rm star}$ is the mass fraction retained in long-lived stars after stellar mass loss. Furthermore, \citet{lilly13_bathtube} pointed out that the difference in $\beta$ between the sAR and the sSFR could be a result of the curvature in the $f_{\rm star}(M_{\star})$ relation that can itself be traced to the curvature in the mass-metallicity relation. Nevertheless, the mass dependence is rather weak either way, and it has only a secondary effect on our current study.

\subsection{Scatter of the MS}

Observations typically reveal a MS scatter of $\sigma_{\mathrm{MS}}\simeq0.3~\mathrm{dex}$ \citep[e.g., ][]{noeske07, rodighiero11, whitaker12, schreiber15}. Clearly, the measured scatter depends on the exact selection criteria for the SFGs as well as on the uncertainties in the stellar mass and SFR indicators. 

As discussed by \citet{speagle14}, each MS observation is measured within a predefined redshift window. Because of this, the observed scatter about the MS, $\sigma_{\mathrm{MS}}$, is actually the intrinsic scatter about the MS convolved with the MS's evolution within the given time interval. The $\sigma_{\mathrm{MS}}$ measured from the observations are therefore overestimates. \citet{speagle14} deconvolved the observed scatter by using their best fits to approximate the MS evolution within each time interval, and subtracted this evolution from the observed scatter. Furthermore, the observation-induced scatter is also taken into account. They find that the true, observation-corrected scatter is estimated to be $\sim0.2-0.3~\mathrm{dex}$, respectively. They find the scatter to be roughly constant with cosmic time. 

In the simulations, we can directly measure the true scatter about the MS. We measure a scatter of $\sigma_{\mathrm{MS}}=0.27\pm0.01~\mathrm{dex}$ for $z=3-6$ (error obtained from bootstrapping), i.e., consistent with the observational estimates. We find a weak trend with cosmic time: the scatter increases from $0.25_{-0.01}^{+0.02}~\mathrm{dex}$ at $z\sim5$ to $0.31\pm0.01~\mathrm{dex}$ at $z\sim3$. This may reflect more contamination by quenching galaxies at lower redshifts, leading to a bend of the MS downwards at the massive end.

%%%%%%%%%%%%%%%%%%%%%%%%%%%%%%%%%
\section{Galaxy Properties across the MS}\label{sec:GalaxyProperties}

In this section, we measure galaxy properties in the simulations as they evolve along and across the MS. We attempt to correlate the galaxy migration above and below the MS ridge with the major events of compaction, depletion, and quenching during the galaxy's evolution. We first look at a few galaxies individually. Afterwards, we determine gradients across the MS from all galaxies in the simulations and compare them with recent observations by \citet{genzel15}. Finally, we investigate the timescale for the oscillation along the MS ridge by conducting a Fourier analysis.

\subsection{Individual Galaxies}\label{subsec:MSgrad_individual}

\begin{figure*}
\includegraphics[width=\textwidth]{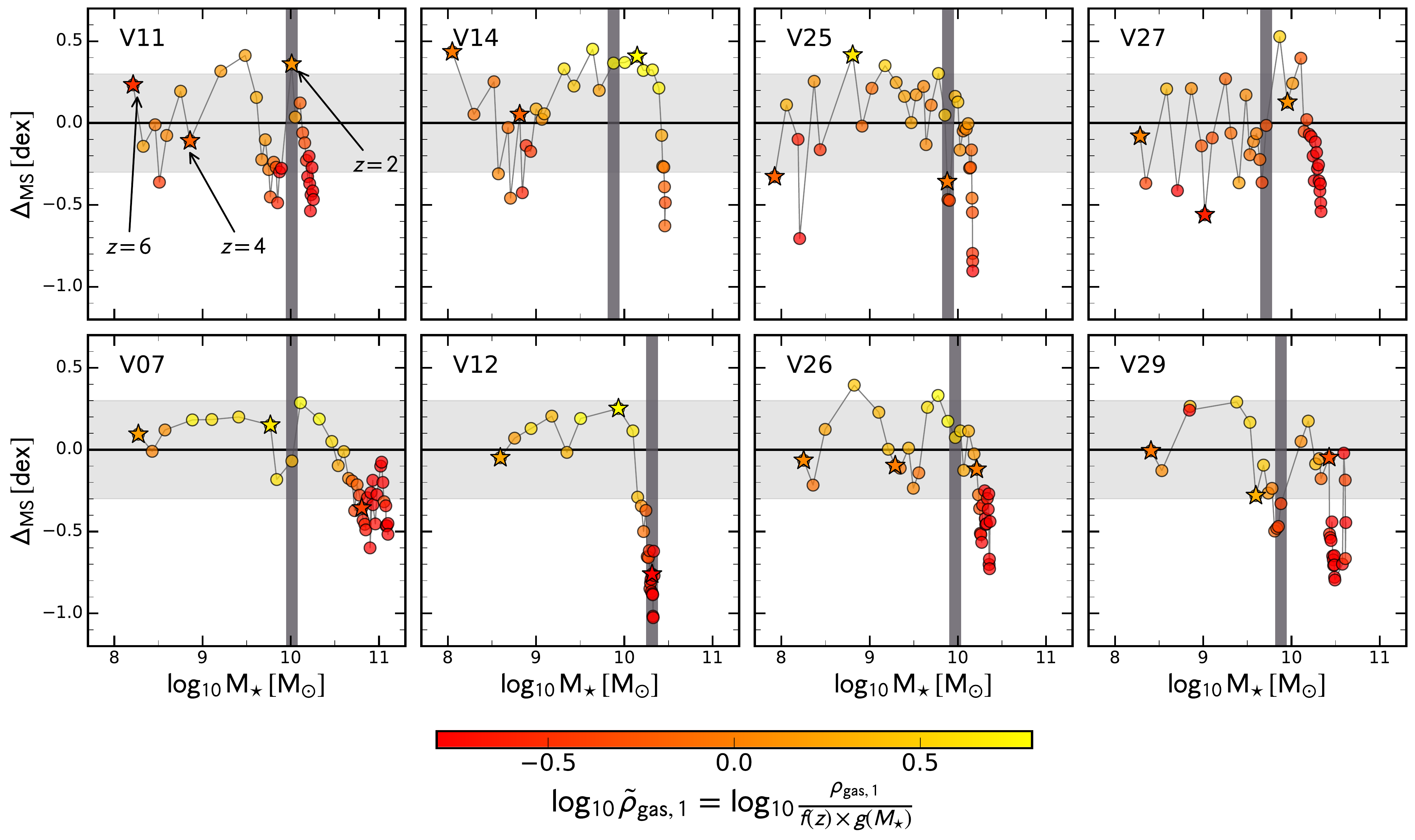} 
\caption{Evolution of galaxy properties along and across the MS for eight simulated galaxies. Each panel shows the distance from the MS $\Delta_{\mathrm{MS}}=\log_{10}\left(\mathrm{sSFR}/\mathrm{sSFR}_{\mathrm{MS}}\right)$ as a function of $M_{\star}$. The number in the upper left corner shows the identification number of the galaxy (see Table~\ref{tab:sample}). The upper panels show four examples of low-mass galaxies, while the lower panels show examples of high-mass galaxies. The stars mark the redshifts $z=6, 4,$ and 2 from left to right. The colour coding corresponds to the gas density within the central 1 kpc, $\rho_{\rm gas,1}$, corrected for the systematic dependence on $z$ and $M_{\star}$ using $f(z)=\xi+\zeta\times\log_{10}(1+z)$ and $g(M_{\star})=\eta+\gamma\times(\log_{10}(M_{\star})-10.5)$, see Section~\ref{subsec:MS_gradients} and Appendix~\ref{App:gradient}. The grey vertical line indicates the stellar mass at the time when the galaxy's halo mass has reached $M_{\rm vir}=3\times10^{11}~M_{\odot}$, above which quenching is expected to be more likely. Galaxies at the top of the MS have about an order of magnitude higher $\rho_{\rm gas,1}$ than at the bottom of the MS.}
\label{Fig:Delta_MS_gas_1kpc}
\end{figure*}

\begin{figure*}
\includegraphics[width=\textwidth]{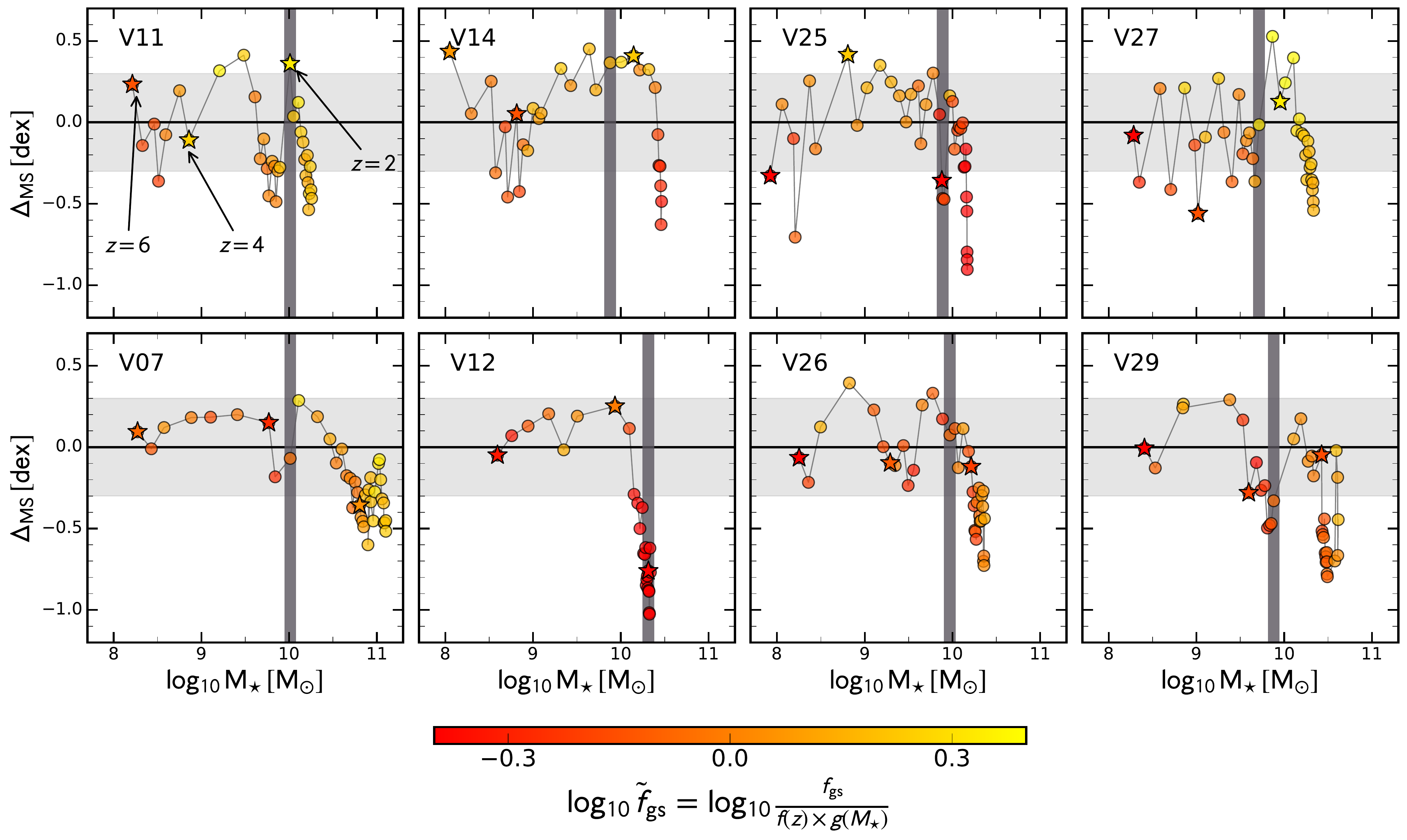} 
\caption{Same as Figure~\ref{Fig:Delta_MS_gas_1kpc}, but the colour coding indicates the gas to stellar mass ratio $f_{\rm gs}$ corrected to take out the overall $z$-evolution and $M_{\star}$-dependence ($f(z)$ and $g(M_{\star})$). Galaxies at the top of MS tend to be more gas-rich, whereas galaxies at the bottom are typically gas-poor.}
\label{Fig:Delta_MS_fgas}
\end{figure*}

\begin{figure*}
\includegraphics[width=\textwidth]{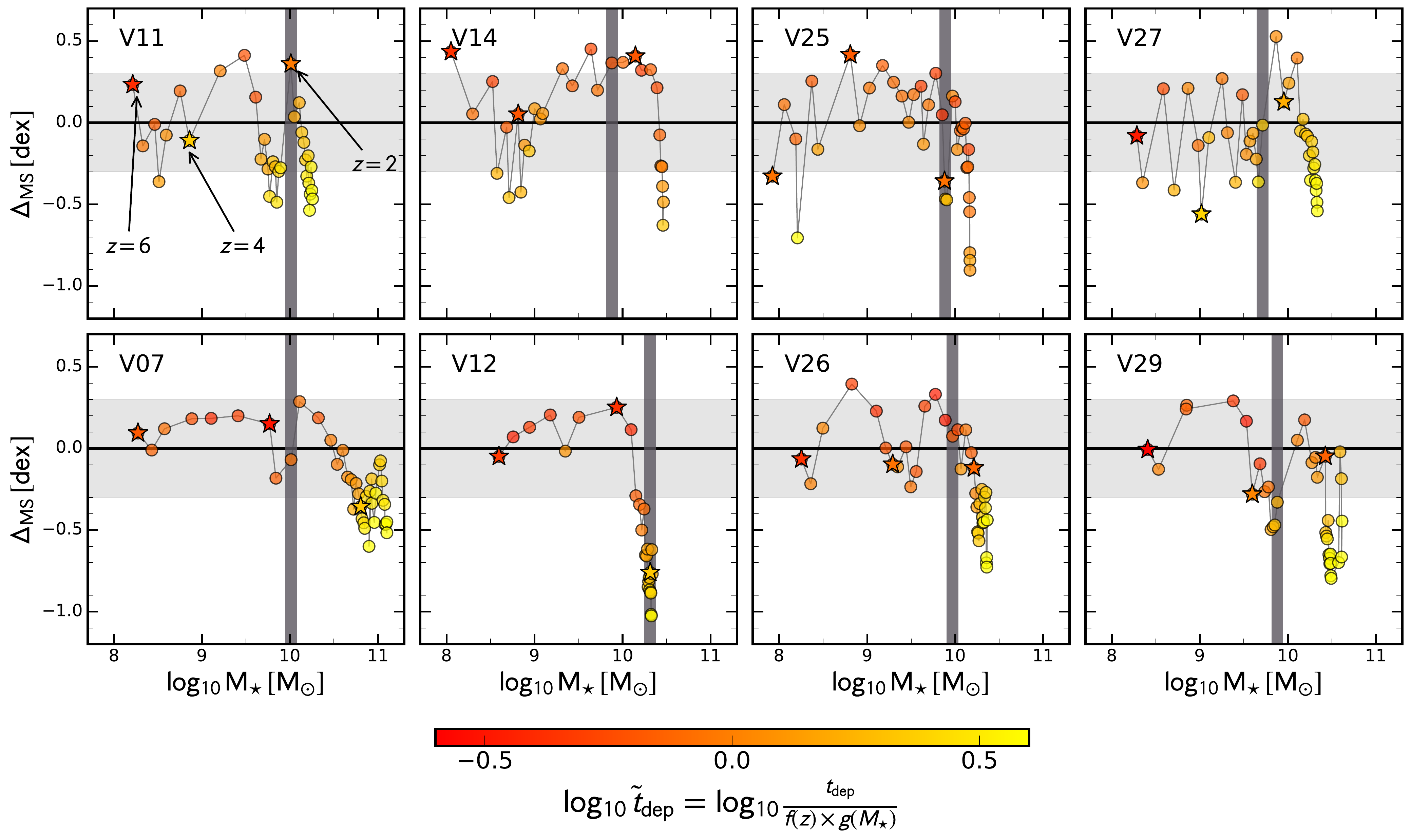} 
\caption{Same as Figure~\ref{Fig:Delta_MS_gas_1kpc}, but the symbol size indicates the depletion time ($t_{\mathrm{dep}}=M_{\rm gas}/\mathrm{SFR}$) corrected to take out the overall $z$-evolution and $M_{\star}$-dependence ($f(z)$ and $g(M_{\star})$). Galaxies at below the MS ridge tend to have longer depletion times than galaxies above it.}
\label{Fig:Delta_MS_tdep}
\end{figure*}

In Figures~\ref{Fig:Delta_MS_gas_1kpc}, \ref{Fig:Delta_MS_fgas}, and \ref{Fig:Delta_MS_tdep}, we show the evolution of gas density within the central 1 kpc ($\rho_{\rm gas,1}$), the overall gas to stellar mass ratio within the galaxy ($f_{\rm gs}=M_{\rm gas}/M_{\star}$), and the depletion time ($t_{\rm dep}$; Equation~\ref{eq:tdep_def}), for eight galaxies along the MS. In the three figures, we plot the distance from the MS, $\Delta_{\mathrm{MS}}=\log_{10}\left(\mathrm{sSFR}/\mathrm{sSFR}_{\mathrm{MS}}\right)$, as a function of stellar mass, $M_{\star}$. The grey vertical line indicates the stellar mass when the galaxy's halo reaches $M_{\rm vir}=10^{11.5}~M_{\odot}$, argued to mark the threshold halo mass for virial shock heating and thus long-term quenching \citep{dekel06, zolotov15}.

Since we wish to quantify the gradients of $\rho_{\rm gas,1}$, $f_{\rm gs}$, and $t_{\mathrm{dep}}$ across the universal MS, we have to correct for their intrinsic cosmic time evolution. For example, we saw in Section~\ref{sec:GasContent} that the gas content of a galaxy is a strong function of redshift and stellar mass. Since we follow a galaxy sample through cosmic time, the time evolution is naturally associated with growth of stellar mass. We quantify this evolution with cosmic time and stellar mass in detail in Section~\ref{subsec:MS_gradients} and in Appendix~\ref{App:gradient}. Briefly, we determine the average evolution by the functions $f(z)=\xi+\zeta\times\log_{10}(1+z)$ and $g(M_{\star})=\eta+\gamma\times(\log_{10}(M_{\star})-10.5)$ deduced for the quantity of interest from all galaxies that lie close to the MS ridge ($|\Delta_{\mathrm{MS}}|<0.15$), assuming that $f(z)$ and $g(M_{\star})$ are independent from each other. The best-fitting parameters can be found in Table~\ref{tab:best_fit_gradients}. We then divide all individual measurements by $f(z)$ and $g(M_{\star})$. 

The eight example galaxies shown in Figures~\ref{Fig:Delta_MS_gas_1kpc}, \ref{Fig:Delta_MS_fgas}, and \ref{Fig:Delta_MS_tdep} are prototypical for two mass bins at $z=2$: the upper four galaxies (11, 14, 25, 27) belong to the low-mass bin ($\log_{10}~M_{\star}/M_{\odot}<10.2$), while the lower four (07, 12, 26, 29) are from the high-mass bin ($\log_{10}~M_{\star}/M_{\odot}>10.2$). We caution that the galaxies which are in a certain mass bin at $z=2$ are not necessarily in the same mass bin at all times. We make this division in order to differentiate more evolved versus less evolved galaxies at a certain epoch, motivated by the finding of \citet{zolotov15} that more massive galaxies at $z=2$ tend to quench earlier and more decisively. 

The first insight from these plots is that the individual galaxies tend not to be super-MS or sub-MS at all times; they evolve through super-MS and sub-MS phases. Most galaxies actually oscillate around the MS on timescales of $\sim0.4~t_{\rm H}$, where $t_{\rm H}$ is the Hubble time at that epoch. This corresponds to $\sim0.6-1.3~\mathrm{Gyr}$ at $z=4-2$. The more evolved galaxies, of higher masses at a given redshift, show fewer oscillations, one or two major compaction events followed by a decisive quenching. We investigate this further by performing a Fourier analysis in Section~\ref{subsec:Oscillation}. Furthermore, we explore in Appendix~\ref{App:thin} how the results based on a higher temporal resolution of snapshots compare with the results based on our standard resolution in Appendix~\ref{App:thin}. In Figure~\ref{FigApp:thin}, we plot the evolution of six simulated galaxies (11, 12, 14, 25, 26, and 27) along the MS with high (upper panels) and standard (lower panels) temporal resolution. This figure confirms that the standard temporal resolution (steps of 100 Myr) traces the main features as well as the short term fluctuations.

All galaxies evolve through certain sub-MS phases. At high redshifts, when the galaxy lives in a halo of relatively low mass ($M_{\rm vir}<10^{11}~M_{\odot}$), these are only quenching attempts, i.e. the galaxy depletes some of its gas content and remains below the MS ridge for some time, but it quickly regains its position near or even above the MS ridge. This can be explained by fresh gas that flows through the halo into the galaxy, replenishing the gas in the disc, leading to a new compaction event and renewed star formation. Only after the galaxy's halo reaches a critical virial mass of $M_{\rm vir}\ga10^{11.5}~M_{\odot}$, and at late enough redshifts, the quenching process may continue for many 100 Myr up to several Gyr, allowing a successful quenching as opposed to a short-term quenching attempt at high-$z$. 

Focusing first on the colour gradients in Figure~\ref{Fig:Delta_MS_gas_1kpc}, we find that all eight galaxies show a strong positive correlation between $\Delta_{\mathrm{MS}}$ and $\rho_{\rm gas,1}$. Galaxies on the upper envelope of the MS have about an order of magnitude higher central densities than galaxies below the MS, i.e., they are star-forming, compact blue nuggets, in which the central stellar density is soon to reach its peak value. Figure~\ref{Fig:Delta_MS_fgas} shows that galaxies on the upper envelope of the MS tend to have a higher gas to stellar mass ratio than galaxies at the lower envelope. In Figure~\ref{Fig:Delta_MS_tdep}, we see that galaxies at the top of the MS typically have a short $t_{\mathrm{dep}}$ of $\sim300~\mathrm{Myr}$. On the other hand, galaxies at the bottom of the MS tend to have a longer $t_{\mathrm{dep}}$ of up to $\sim1~\mathrm{Gyr}$. Quenching galaxies all have low values of $\rho_{\rm gas,1}$ ($<10^6~M_{\odot}/\mathrm{kpc}^3$), low values of $f_{\rm gs}$ ($<0.1$), and long depletion times $t_{\mathrm{dep}}$ ($>2~\mathrm{Gyr}$).

\subsection{Overall Population} \label{subsec:MS_gradients}
\subsubsection{Gas-related gradients} 

\begin{figure}
\includegraphics[width=\linewidth]{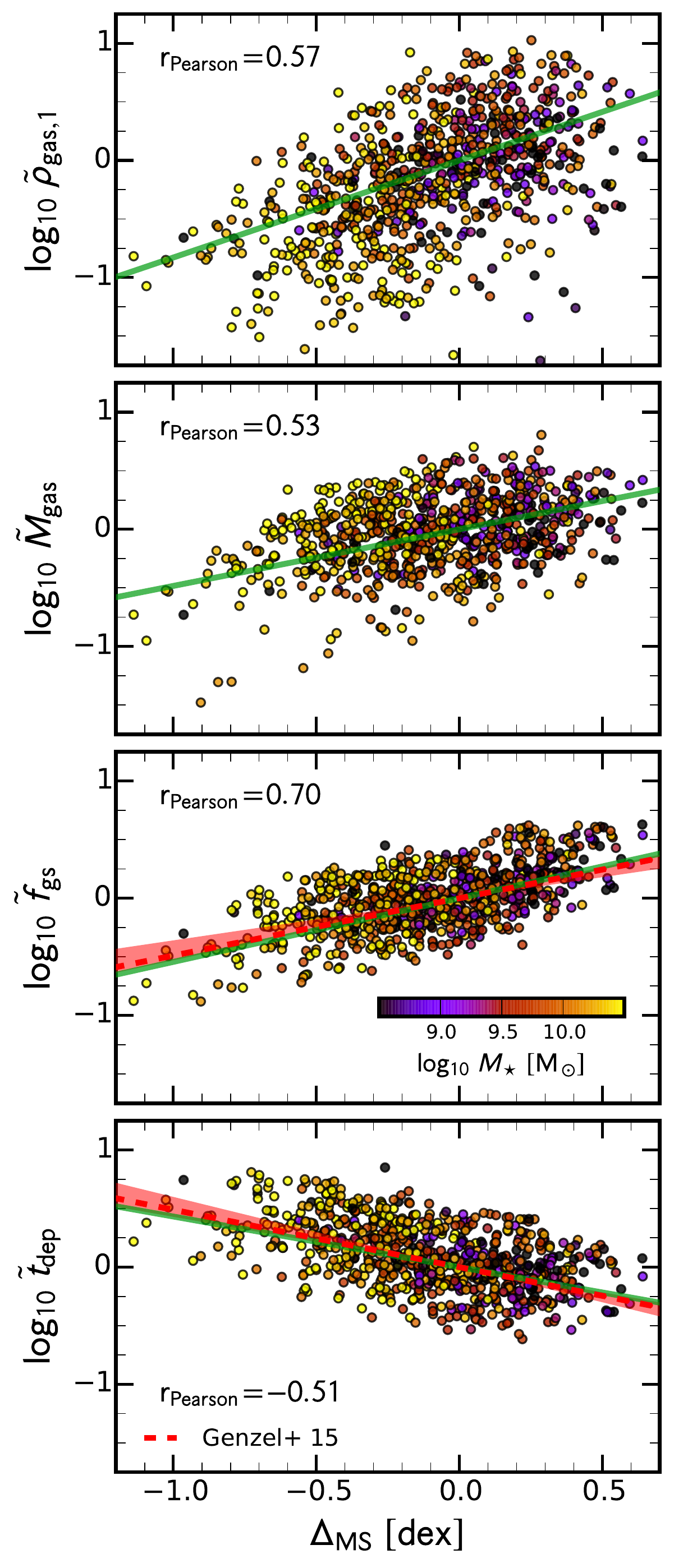} 
\caption{Gas-related galaxy properties across the MS. Shown from top to bottom are the gas mass within 1 kpc, total gas mass, gas mass to stellar mass ratio, and depletion time as a function of the distance from the MS $\Delta_{\mathrm{MS}}$. The quantities are corrected for the systematic dependence on $z$ and $M_{\star}$. Linear regression lines are shown (green lines), and the correlation coefficients are quoted. The color coding corresponds to stellar mass. The red dashed line and the shaded region show the best-fitting and its uncertainty of the observations by \citet{genzel15}, indicating an excellent agreement with the gradients in the simulations. For the uncorrected quantities as well as the systematic dependence on redshift and stellar mass see Figure~\ref{FigApp:gradient}. }
\label{Fig:Gradients_MS}
\end{figure}

\begin{table*}
  \begin{tabular}{lccccccc}
  \hline\hline
      & \multicolumn{2}{c}{$z$-dependence} & \multicolumn{2}{c}{$M_{\star}$-dependence} & \multicolumn{2}{c}{$\Delta_{\mathrm{MS}}$-dependence} & $r_{\rm Pearson}$ \\ \hline
      & \multicolumn{2}{c}{$f(z)=\xi+\zeta\times\log_{10}(1+z)$} & \multicolumn{2}{c}{$g(M_{\star})=\eta+\gamma\times(\log_{10}M_{\star}-10.5)$} & \multicolumn{2}{c}{$Q=\alpha+\delta\times\Delta_{\mathrm{MS}}$} & \\ \hline\hline
    This work & $\xi$ & $\zeta$ & $\eta$ & $\gamma$ & $\alpha$ & $\delta$ & \\ \hline
    $\rho_{\rm gas,1}$ 					 & $+7.28\pm0.12$ & $+0.30\pm0.21$ & $+0.31\pm0.05$ & $+0.36\pm0.04$ & $-0.06\pm0.02$ & $+0.83\pm0.04$ & $0.57\pm0.02$ \\
    $M_{\rm gas}$ 							 & $+10.21\pm0.09$ & $-1.71\pm0.15$ & $+0.28\pm0.03$ & $+0.32\pm0.03$ & $+0.00\pm0.01$ & $+0.48\pm0.03$ & $0.53\pm0.02$ \\
    $f_{\rm gs}$               			     & $-1.31\pm0.06$ & $+1.63\pm0.10$ & $-0.13\pm0.02$ & $-0.15\pm0.02$ & $+0.02\pm0.01$ & $+0.54\pm0.02$ & $0.70\pm0.02$ \\
    $t_{\mathrm{dep}}$                  & $-0.19\pm0.07$ & $-0.39\pm0.11$ & $-0.17\pm0.02$ & $-0.19\pm0.02$ & $+0.02\pm0.01$ & $-0.43\pm0.02$ & $-0.51\pm0.03$ \\  
    $\Sigma_{M_{\star},1}$             & $+10.26\pm0.13$ & $-2.73\pm0.22$ & $+0.43\pm0.04$ & $+0.49\pm0.04$ & $-0.06\pm0.01$ & $-0.12\pm0.04$ & $-0.12\pm0.02$ \\
    $R_{\rm e}$               			     & $+0.70\pm0.06$ & $-1.03\pm0.10$ & $-0.03\pm0.02$ & $-0.04\pm0.02$ & $+0.04\pm0.01$ & $+0.03\pm0.02$ & $0.05\pm0.02$ \\
    $n$               	             		     & $+0.75\pm0.09$ & $-0.57\pm0.14$ & $+0.12\pm0.03$ & $-0.12\pm0.03$ & $+0.17\pm0.01$ & $+0.16\pm0.03$ & $0.17\pm0.02$ \\
    \hline\hline    
    Observations$^{a}$ & & & & \\ \hline
    $f_{\rm gs}$							      & & $+2.68\pm0.05$ & & $-0.37\pm0.04$ & & $+0.49\pm0.03$ & \\
    $t_{\mathrm{dep}}$                   & & $-0.34\pm0.05$ & & $+0.01\pm0.03$ & & $-0.49\pm0.02$ &  \\ \hline\hline 
  \end{tabular}
  \caption{List of best-fitting parameters for the gradients across the MS (Figures~\ref{Fig:Gradients_MS} and \ref{Fig:Gradients_MS_structure}) in comparison with the values of observations. Last column shows the Pearson's correlation coefficient $r_{\rm Pearson}$. \newline
  $^{a}$: Data taken from Table 3 and 4 (global fits) of \citet{genzel15}. }
  \label{tab:best_fit_gradients}
\end{table*}

The trends in Figures~\ref{Fig:Delta_MS_gas_1kpc}, \ref{Fig:Delta_MS_fgas} and \ref{Fig:Delta_MS_tdep} are studied more quantitatively in this section. We extend the focus from a few galaxies to all the 26 simulated galaxies of this study. Figure~\ref{Fig:Gradients_MS} shows from top to bottom the central gas density within 1 kpc ($\rho_{\rm gas,1}$), the total gas mass ($M_{\rm gas}$), gas to stellar mass ratio ($f_{\rm gs}$), and depletion time ($t_{\mathrm{dep}}$), corrected for the $z$-dependence and $M_{\star}$-evolution, as a function of the distance from the MS ridge ($\Delta_{\mathrm{MS}}$) for all galaxies at $z=1-6$. 

We have corrected for the systematic time evolution of galaxies near the MS ridge because these key quantities may also evolve with cosmic time. In the simulations, we follow individual galaxies through cosmic time. This leads to a degeneracy between the redshift and the mass evolution, since all galaxies increase their mass with decreasing redshift. To correct for the cosmic time evolution, we first fit the $z$-evolution of the galaxies that are near the MS ridge, $|\Delta_{\mathrm{MS}}|<0.15$. Varying the threshold for $|\Delta_{\mathrm{MS}}|$ between 0.05 and 0.20 dex has no noticeable effect on the fits. After correcting for the $z$-evolution, we fit the $M_{\star}$-dependence. In this procedure, we assume that the cross-terms between $z$-evolution and $M_{\star}$-dependence are negligible, i.e., we can practically determine them independently of each other. 

In Appendix~\ref{App:gradient}, Figure~\ref{FigApp:gradient}, we show a more extended version of Figure~\ref{Fig:Gradients_MS}. In the left panels, we show the uncorrected quantities. In the middle-left and middle-right panels, we show the $z$-dependence and $M_{\star}$-dependence, respectively. The most right panels show the corrected quantities, i.e., the same as shown in Figure~\ref{Fig:Gradients_MS}. We find a steep redshift evolution ($f(z)=\xi+\zeta\times\log_{10}(1+z)$) for $M_{\rm gas}$ and $f_{\rm gs}$ with $\zeta=-171\pm0.14$ and $+1.63\pm0.17$, respectively. For $\rho_{\rm gas,1}$ and $t_{\mathrm{dep}}$, the $z$-evolution is much shallower with $\zeta=+0.30\pm0.26$ and $-0.39\pm0.19$, respectively. Table~\ref{tab:best_fit_gradients} lists the best-fitting values. Following the $z$-correction, we fit the $M_{\star}$-dependence with the relation $g(M_{\star})=\eta+\gamma\times(\log_{10}(M_{\star})-10.5)$. We find $\gamma=+0.36\pm0.05$, $+0.32\pm0.03$, $-0.15\pm0.03$, and $-0.19\pm0.03$ for $\rho_{\rm gas,1}$, $M_{\rm gas}$, $M_{\rm gas}/M_{\odot}$, and $t_{\mathrm{dep}}$, respectively (Table~\ref{tab:best_fit_gradients}). 

In Figure~\ref{Fig:Gradients_MS}, we fit each of the corrected quantities $Q$ with $Q=\alpha+\delta\times\Delta_{\mathrm{MS}}$, which is shown as a green line. We find $\delta=+0.83\pm0.03$, $\delta=+0.48\pm0.02$, $\delta=+0.54\pm0.02$, and $\delta=-0.43\pm0.02$ for $\rho_{\rm gas,1}$, $M_{\rm gas}$, $f_{\rm gs}$, and $t_{\mathrm{dep}}$, respectively (Table~\ref{tab:best_fit_gradients}). Interestingly, the gradient of $\rho_{\rm gas,1}$ across the MS is the steepest, and the correlation is strongly positive (Pearson's correlation coefficient $r=0.57$). This demonstrates that galaxies above the MS ridge have a significantly higher central gas density than galaxies below it: going from $\Delta_{\rm MS}=0.5$ dex below the MS ridge to 0.5 dex above it, we find that $\rho_{\rm gas,1}$ increases by nearly an order of magnitude. The gradients across the MS for the other quantities are very similar to each other (a slope of $\sim0.5$) with $|r|\sim0.5-0.7$. 

Putting it all together, as seen by comparing Figure~\ref{Fig:Delta_MS_gas_1kpc} to Figures~\ref{Fig:Delta_MS_fgas} and \ref{Fig:Delta_MS_tdep}, the gradients across the MS are intimately related to the galaxy evolutionary phases of compaction to a blue nugget followed by central gas depletion and quenching. Galaxies above the MS are compact, gas-rich systems with high SFRs and short depletion times, whereas galaxies below are either more diffuse or depleted from central gas, of lower SFR and longer depletion times.

\subsubsection{Stellar-structure gradients} 

\begin{figure}
\includegraphics[width=\linewidth]{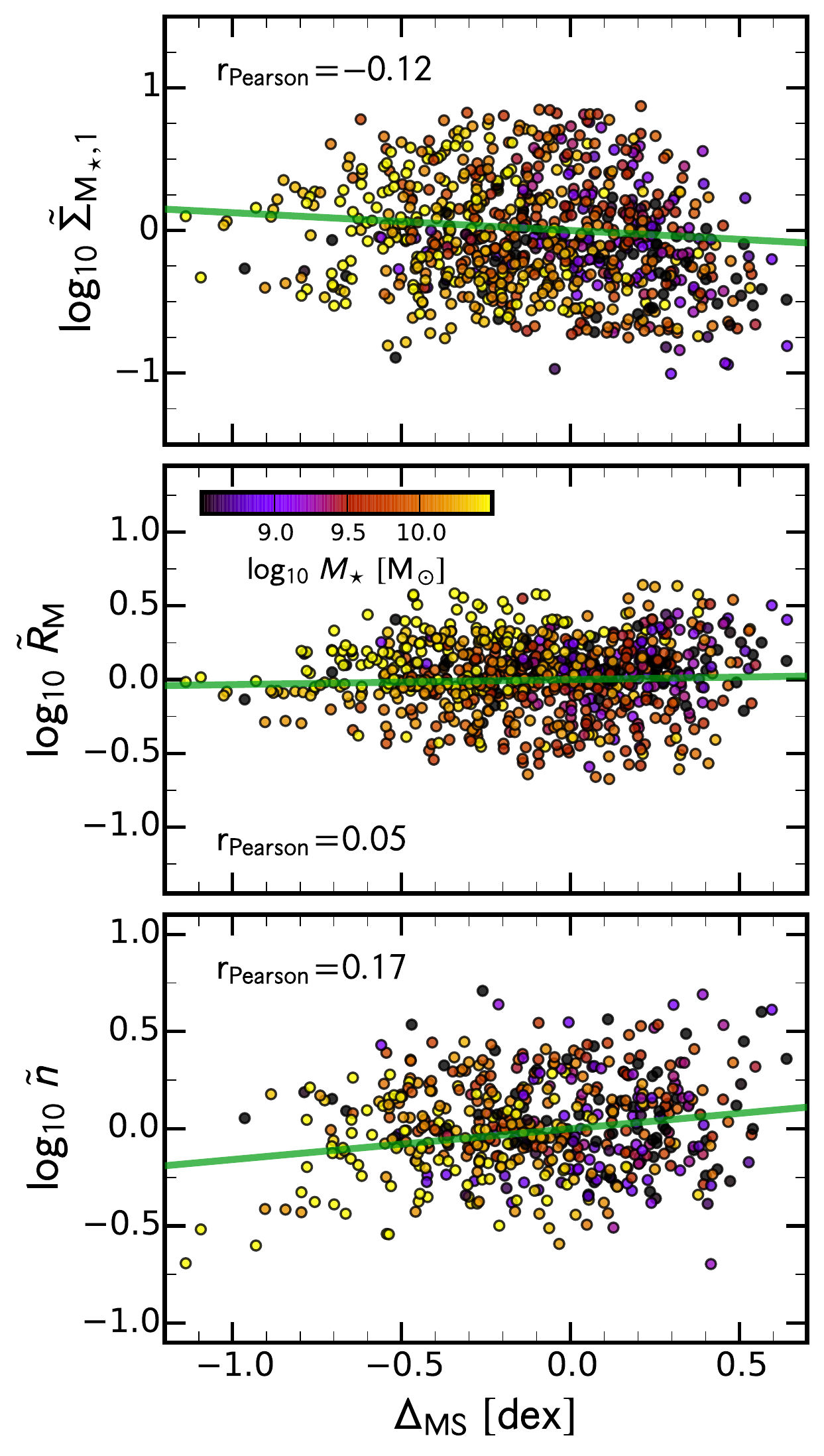} 
\caption{ Stellar-structure properties across the MS. Shown from top to bottom are the surface mass density within 1 kpc ($\Sigma_{M_{\star},1}$), the half-mass radius ($R_{\rm e}$), and the S\'{e}rsic index ($n$), corrected for their systematic dependence on redshift and stellar mass, versus the deviation from the MS ridge ($\Delta_{\rm MS}$). Linear regression lines are shown (green lines), and the correlation coefficients are quoted. The color coding corresponds to stellar mass. We see no significant systematic gradients across the MS. For the uncorrected quantities as well as the systematic dependence on redshift and stellar mass see Figure~\ref{FigApp:gradient_structure}. }
\label{Fig:Gradients_MS_structure}
\end{figure}

In this section, we investigate stellar-structure gradients across the MS. We follow the approach used before for the gas-related gradients. Figure~\ref{Fig:Gradients_MS_structure} shows from top to bottom the central stellar mass surface density within 1 kpc ($\Sigma_{M_{\star},1}$), the half-mass radius ($R_{\rm e}$), and the S\'{e}rsic index ($n$), corrected for their systematic variations with redshift and with stellar mass, as a function of distance from the MS ridge ($\Delta_{\rm MS}$) for all galaxies and snapshots at $z=6-1$. As before, we show in Appendix~\ref{App:gradient}, Figure~\ref{FigApp:gradient_structure}, a more extended version of Figure~\ref{Fig:Gradients_MS_structure}, including the uncorrected quantities, and how we gradually account for the $z$-evolution and then the $M_{\star}$-dependence. The best-fitting values for the dependences on $z$ and on $M_{\star}$, as well as the gradients across the MS are listed in Table~\ref{tab:best_fit_gradients}. 

In Figure~\ref{Fig:Gradients_MS_structure}, the green lines show the linear regressions of the corrected quantities. They are much flatter in comparison with the gas-related gradients from above. We find $\delta=-0.12\pm0.04$, $\delta=+0.03\pm0.02$, and $\delta=+0.16\pm0.03$ for $\Sigma_{M_{\star},1}$, $R_{\rm e}$, and $n$, respectively. Furthermore, all three stellar structure quantities are not significantly correlated with the distance from the MS: the Pearson's correlation coefficient is $|r|<0.20$. 

\subsection{Gradients across the MS in Observations} \label{subsec:MS_gradients_OBS}

\citet[][hereafter G15]{genzel15} present CO-based and Herschel dust-based scaling relations of $t_{\mathrm{dep}}$ and of $f_{\rm gs}$ as a function of redshift, $\Delta_{\rm MS}$ and $M_{\star}$, for each of $\sim500$ SFGs between $z\sim0$ and 3. The best fit relations are spelled out in Table~\ref{tab:best_fit_gradients}. 

Focusing first on $f_{\rm gs}$, we find a very similar scaling with $z$ and $M_{\star}$ in the simulations: G15 find as best fit for the $z$-dependence, combining CO and dust data\footnote{For the global combined CO+dust fit, G15 first added 0.1 dex to all CO values, and likewise subtracted 0.1 dex for all dust values before carrying out the global fit, in order to bring the two data sets to the same zero point. The values which we quote in the brackets are for the individual fits to the CO and dust data.}, $\zeta=+2.68\pm0.05$ ($+2.71$ for the CO data and $+2.32$ for the dust data, respectively), which is slightly larger than our value of $\zeta=+1.63\pm0.10$. For the $M_{\star}$-dependence, G15 find $\gamma=-0.37\pm0.04$ ($-0.35$ for the CO data and $-0.40$ for the dust data, respectively), also slightly larger than our value of $\gamma=-0.15\pm0.02$. However, despite these different scalings with $z$ and $M_\star$, the obtained gradient across the MS is in very good agreement: G15 determined the gradient of $f_{\rm gs}$ to be $\delta=0.49\pm0.03$ ($0.53$ for the CO data and $0.36$ for the dust data, respectively), while we measure a value of $\delta=0.54\pm0.02$. In Figure~\ref{Fig:Gradients_MS}, the best-fitting line of G15 and its uncertainty are indicated as a red dashed line and as a red shaded area. 

Focusing now on $t_{\mathrm{dep}}$, we find an even better agreement between the $z$-dependence and $M_{\star}$-dependence of G15 and ours. For the $z$-dependence, G15 find $\zeta=-0.34\pm0.05$ ($-0.20$ CO, $-0.77$ dust), while we find $\zeta=-0.39\pm0.11$. For the $M_{\star}$-dependence, G15 find $\gamma=+0.01\pm0.03$ ($-0.01$, $0.00$), while we find $\gamma=-0.19\pm0.02$. For the $t_{\mathrm{dep}}$ gradient across the MS, G15 determines $\delta=-0.49\pm0.02$ ($-0.43$, $-0.59$), while we find $\delta=-0.43\pm0.02$, i.e., our values are consistent within the systematic uncertainty. 

\citet{wuyts11} analyse the dependence of galaxy structure (size and S\'{e}rsic index) on the position of the galaxies with respect to the MS ridge at $z=0-2$. They performed the structural measurements on the light at the longest wavelength, high-resolution imaging, i.e., on $z_{850}$, $I_{814}$, and $H_{160}$, which are rest-frame UV to optical at $z\sim2$. At $z\sim2$, they find no significant gradient of galaxy size across the MS ($\Delta_{\rm MS}=-1.0$ to $1.0$). When galaxies lie more than two orders of magnitudes below the MS, i.e., quiescent galaxies, the sizes decrease by $\sim0.2$ dex. The S\'{e}rsic index also tends to be roughly the same, $n\sim1$, between $\Delta_{\rm MS}=-1.0$ and $1.0$, i.e., there is no significant gradient of $n$ across the MS. Again, when the galaxies are two orders of magnitude below the MS ridge, the S\'{e}rsic index increases to $n\approx4.0$. Our simulations show the same null gradients in size and S\'{e}rsic index about the MS ridge. As shown in \citet{tacchella15_profile}, for simulated galaxies that evolve along the MS ridge, the S\'{e}rsic index increases from $n\sim1-2$ at early times and low stellar masses ($z\la2.5$, $M_{\star}<10^{10}~M_{\odot}$) to $n\sim4$ at later times and higher masses. Massive galaxies that leave the MS have typically a high S\'{e}rsic index of $n\sim4$. In comparison to observations, where also the most massive galaxies on the MS have $n\sim2-3$, it is important to consider that these measurements were performed on the UV and optical light, i.e., not on the mass as done in the simulations. Light-based profiles are indeed shallower, i.e., have a lower S\'{e}rsic index, as the mass-based profiles, largely due to star-forming clumps in in the outskirts \citep{carollo14, tacchella15_sci}. Furthermore, since we do not trace many galaxies well below the MS, we cannot address here totally quenched galaxies.

Overall, we find excellent agreement between observations and our simulations for the gradients across the MS, despite the different gas fraction estimates for a given $M_{\star}$ (as discussed in Section~\ref{sec:GasContent}). This indicates that the overall trends across the MS are robust, and that the intimate connection with the evolution through compaction, depletion, and quenching, and through phases of blue nuggets, is qualitatively solid.

\subsection{Driver of the MS Gradients } \label{subsec:Driver}

\begin{figure}
\includegraphics[width=\linewidth]{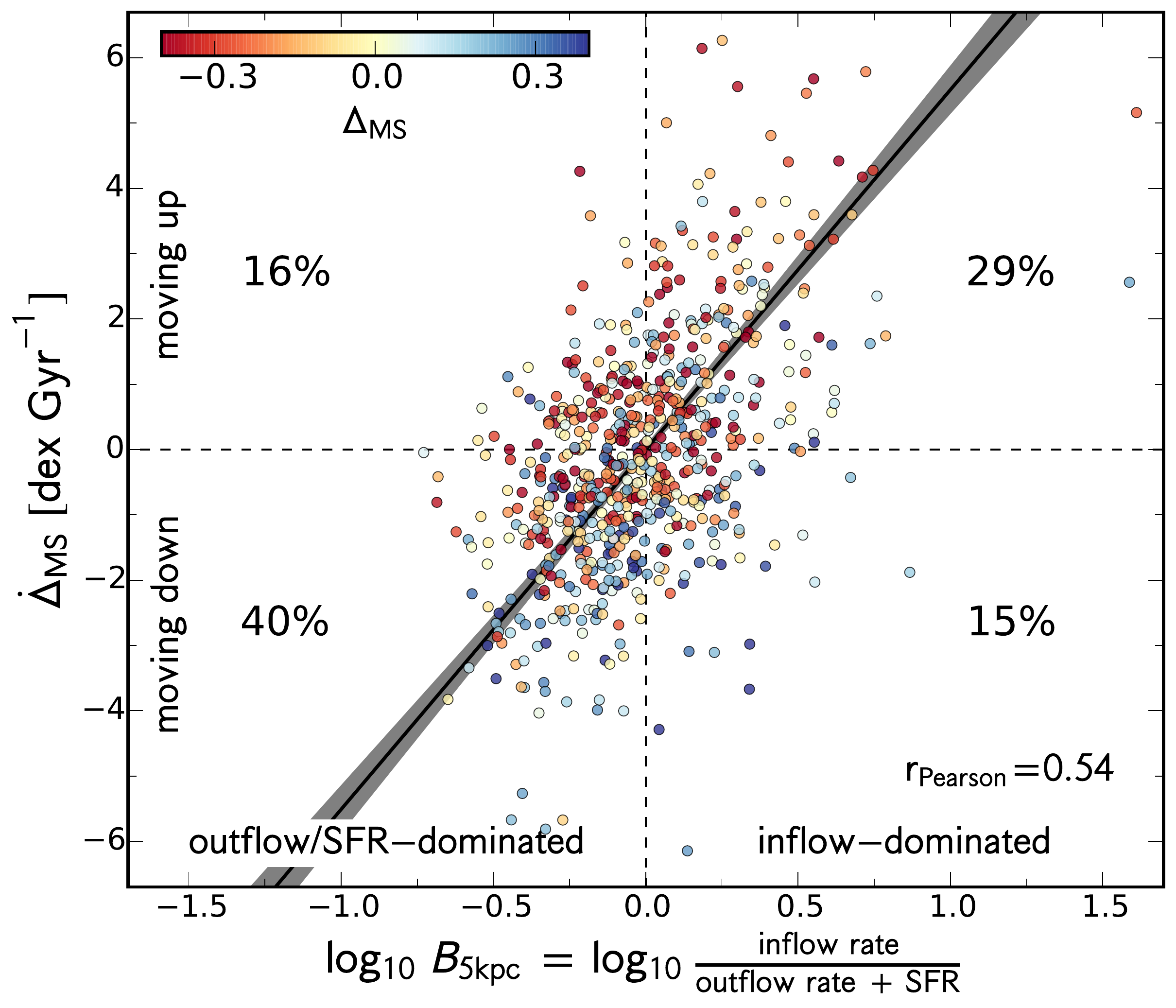} 
\caption{ In the main, large panel, we plot the rate of change of the distance from the MS, $\dot{\Delta}_{\rm MS}$, as a function of the balance of gas input and drainage within 5 kpc, $\log_{10}(B_{5\mathrm{kpc}})=\log_{10}(\mathrm{inflow~rate}/(\mathrm{SFR}+\mathrm{outflow~rate}))$. The colour coding corresponds to the position on the MS, $\Delta_{\rm MS}$. We see a significant correlation between $\dot{\Delta}_{\rm MS}$ and $\log_{10}~B_{5\mathrm{kpc}}$ with $\mathrm{r}_{\rm Pearson}=0.54$. The solid line shows the best fit with $\dot{\Delta}_{\rm MS} = (0.00\pm0.07) + (5.5\pm0.3) \times \log_{10}(B_{5\mathrm{kpc}})$. This shows that galaxies that are moving up towards the upper edge of the MS are inflow-dominated (compaction phase) whereas galaxies that are moving down towards the lower edge of the MS are depleted due to star formation and outflows while the inflow is suppressed (central quenching phase). In Figure~\ref{FigApp:BalanceMS} in Appendix~\ref{App:driver} we split the balance term $\log_{10}(B_{5\mathrm{kpc}})$ into inflow rate, outflow rate, and SFR. }
\label{Fig:Balance}
\end{figure}

To understand the aforementioned gradients across the MS, we study the gas flow in the galaxies in more detail. In particular, we focus on the balance between the gas inflow rate (input term) and SFR plus gas outflow rate (drainage term). We expect galaxies that move upward on the MS, i.e., towards the blue nugget phase, are in an episode of compaction, where the central gas mass increases due to a high inflow rate and low outflow rate. On the other hand, in the post-compaction phase, we expect the opposite, namely, that the inflow rate is outweighed by SFR plus outflow rate. 

We investigate in Figure~\ref{Fig:Balance} the relation between the rate of change of distance from the MS, $\dot{\Delta}_{\rm MS}$, and the balance between the input term and drainage terms of gas in the central 5 kpc, namely $\log_{10}~B_{5\mathrm{kpc}}=\log_{10}[\mathrm{inflow~rate}/(\mathrm{SFR}+\mathrm{outflow~rate})]$. Varying the radius considered between 3 and 10 kpc does not change the result significantly. We have chosen 5 kpc as our fiducial scale since it best captures what happens within the galaxies towards their centres. A too large radius would only capture the gas exchange between the halo and the galaxy, and a too small radius would not capture whole extend of the inflow towards the galaxies' centres. In \citet{tacchella15_profile}, where we focus on the evolution of the surface density profiles, we find that the gas cusp of the compaction phase can reach out to $2-3$ kpc. We have therefore chosen a slightly larger radius. Figure~\ref{FigApp:BalanceMS} in Appendix~\ref{App:driver} we split the balance term $\log_{10}(B_{5\mathrm{kpc}})$ into inflow rate, outflow rate, and SFR, each measured within the central 5 kpc. 

During $\sim70\%$ of the time, this quantity is close to 0 ($|\log_{10}~B_{5\mathrm{kpc}}|<0.3$), i.e., $\mathrm{inflow~rate} \approx \mathrm{SFR}+\mathrm{outflow~rate}$. However, there are episodes (lasting about 20\% of the time) which are inflow-dominated ($\log_{10}~B_{5\mathrm{kpc}}\ga0.3~\mathrm{dex}$). The central gas density $\rho_{\rm gas,1}$ increases quickly ($<300~\mathrm{Myr}$) from $10^7~M_{\odot}$ to a few times $10^8~M_{\odot}$, i.e., this corresponds to dissipative, quick compaction phases of the galaxy gas into a compact, star-forming blue nugget \citep{zolotov15}. \citet{dekel14_nugget} addressed the formation of blue nuggets by wet compaction using as an example the contraction associated with VDI. They applied the requirement that for the inflow to be dissipative and therefore intense, the characteristic timescale for star formation should be longer than the timescale for inflow, namely corresponding to $\log_{10}~B_{5\mathrm{kpc}}>0$. Otherwise, most of the disc mass will turn into stars before it reaches the bulge, the inflow rate will be suppressed, and the galaxy will become an extended stellar system.

In the inflow-dominated phase, when the inflow rate is outweighing the SFR plus outflow rate in the centre, the rate of change of the distance from the MS, $\dot{\Delta}_{\rm MS}$, is positive, i.e., galaxies are moving up in the $\Delta_{\rm MS}-M_{\star}$ plane of the universal MS (Figure~\ref{Fig:Main_Sequence}). Most of these galaxies are below or just on the MS ($\Delta_{\rm MS} \la 0$). There are a few snapshots ($<3\%$) which are inflow-dominated, but where $\dot{\Delta}_{\rm MS}<0$. In these cases, the galaxies are on the upper envelope of the MS ($\Delta_{\rm MS} \sim 0.3$), i.e., they have reached the peak in the universal MS plane and are starting to move downwards. 

The inflow-dominated phases are followed by phases where $\log_{10}~B_{5\mathrm{kpc}} \la -0.3$ dex, where the high SFR and the strong outflows, driven by the high SFR and stellar feedback, outweigh the inflow. This sudden suppression of the inflow at the end of the compaction process happens when the gas disc has shrunk and has not yet been replenished. This is the onset of a central depletion and therefore quenching phase, where the galaxies fall below the MS ridge ($\dot{\Delta}_{\rm MS}<0$). As mentioned before, in low mass haloes at sufficiently high redshifts, these are only quenching attempts, since gas quickly replenishes the disc. This gas is then available to be triggered into a new episode of compaction and high SFR, which causes a subsequent quenching event, and so on. Full quenching of up to several Gyr into low sSFR significantly below the MS can be achieved preferentially at late redshifts, when the replenishment time is longer than the depletion time, and in particular after the galaxy's halo reaches a critical virial mass of $M_{\rm vir}\ga10^{11.5}~M_{\odot}$, corresponding to a stellar mass of $\sim 10^{9.5}~M_{\odot}$.. 

Overall, we find that the gas input and drainage within 5 kpc is strongly correlated with the rate of change of the distance from the MS ($r_{\rm Pearson}=0.54$). From Figure~\ref{FigApp:BalanceMS} in Appendix~\ref{App:driver}, we see that the individual rates are correlated to a lesser degree than the combined balance term $B_{5\mathrm{kpc}}$. The outflow rate and SFR are moderately correlated with $\dot{\Delta}_{\rm MS}$ ($r_{\rm Pearson}=-0.32$), while the inflow rate is not correlated with $\dot{\Delta}_{\rm MS}$.

\subsection{Oscillation Timescale} \label{subsec:Oscillation}

\begin{figure}
\includegraphics[width=\linewidth]{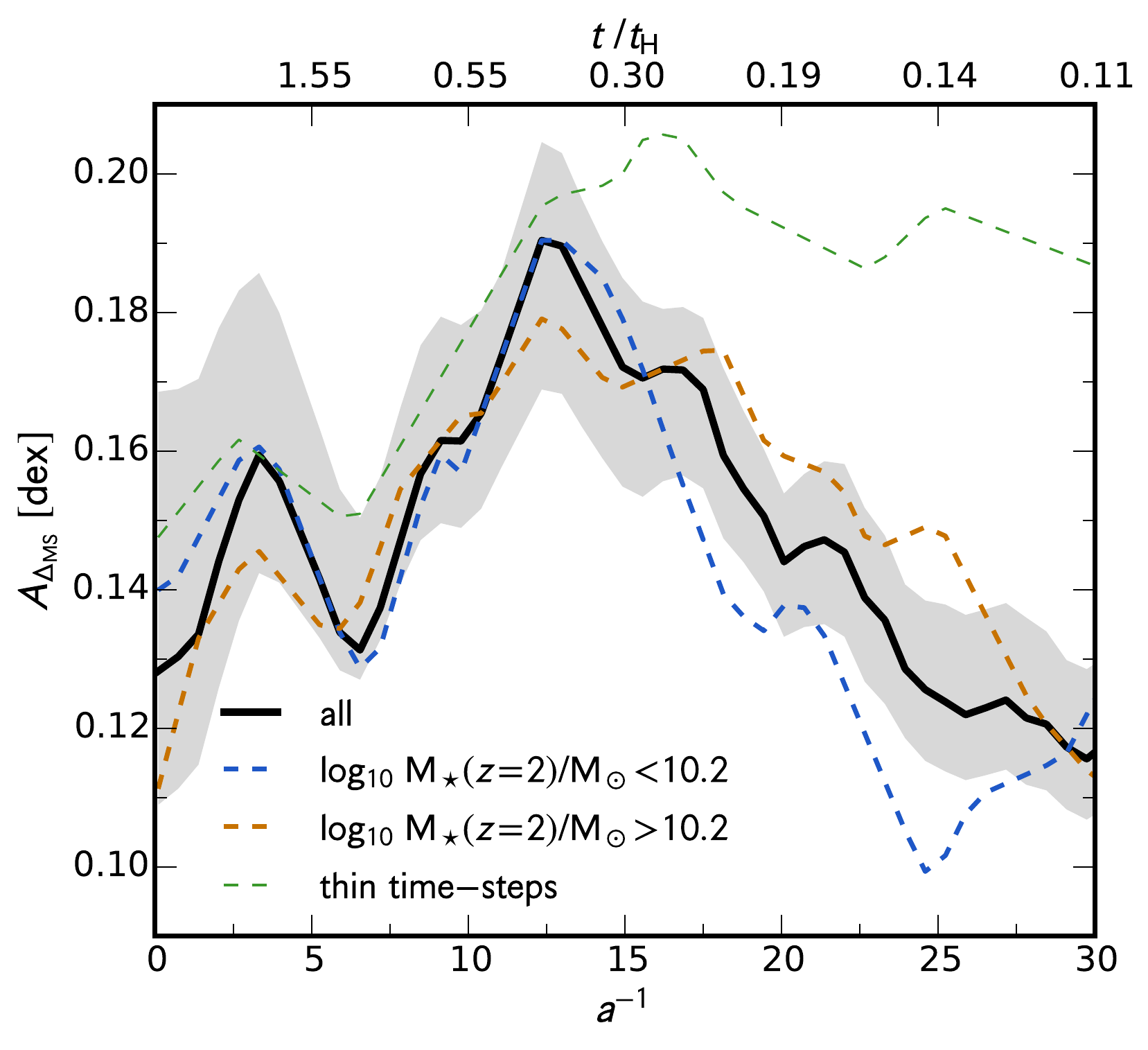} 
\caption{Fourier spectrum of the time for oscillations about the MS. The black line shows the median Fourier spectrum of all simulated galaxies and the shaded area marks the 16 and 84 percentiles. The blue and orange dashed lines refer to the low-mass and high-mass subsamples. The green dashed line refers to the six galaxies for which we have thin time-steps available. The x-axis at the bottom is in units of inverse scale factor $a^{-1}$, whereas the top axis is in units of cosmic time. The broad, global peak indicates that the timescale for the dominant oscillation is $\sim0.2-0.5~t_{\rm H}$. The peak for the low-mass galaxies at $0.43~t_{\rm H}$ is narrower. }
\label{Fig:Fourer_Analysis}
\end{figure}

As discussed before, galaxies oscillate about the MS ridge. To estimate the characteristic timescales for these oscillations more quantitatively, we define the Fourier transform of the displacement from the MS ridge for each of the galaxies individually:
\begin{equation}
A_{\Delta_{{\rm MS}_k}} = \frac{2}{n}\sum_{m=0}^{n-1} \Delta_{{\rm MS}_m} \exp \left( -2\pi i \frac{m k}{n} \right),
\end{equation}
\noindent
where $n$ is the total number of snapshots available for a given galaxy in the redshift range $z=7-3$.

The median of the resulting Fourier spectrum over all galaxies is shown in Figure~\ref{Fig:Fourer_Analysis} as a black solid line. Also shown as dashed-blue and orange lines the are low- and high-mass galaxies, respectively. We find that the main period is $\sim0.2-0.5~t_{\rm H}$ for a full oscillation for galaxies evolving along the MS. The low-mass galaxies show a clear, narrow peak at $\sim0.43~t_{\rm H}$. This corresponds to $\sim0.4~\mathrm{Gyr}$ at $z=6$, and to $\sim0.9~\mathrm{Gyr}$ at $z=3$. The spectrum for the high-mass galaxies is broader with a robust peak in the range $0.2-0.5~t_{\rm H}$.

The green dashed line in Figure~\ref{Fig:Fourer_Analysis} shows the Fourier spectrum for the high temporal resolution. We see that there is more power on small timescale fluctuations, as expected, but the peak at $\sim0.2-0.5~t_{\rm H}$ is confirmed. This is much longer than the dynamical time of the galaxy which is of the order of 30 Myr, i.e., the star-formation recipe or the feedback prescription used in these simulations do not drive the evolution about the MS ridge.

%%%%%%%%%%%%%%%%%%%%%%%%%%%%%%%%%
\section{The Origin of Confinement of the Main Sequence}\label{sec:Confinment}

In this section, we explore the mechanisms responsible for confining the MS to a narrow strip about the MS ridge, connecting the characteristic evolution pattern of high-$z$ galaxies with the evolution along the MS. We apply a toy-model understanding for estimating the timescales that are encoded into the MS.

\subsection{Galaxy Properties across the universal MS}

We have seen in Sections~\ref{subsec:MSgrad_individual} and \ref{subsec:MS_gradients} that differences in the position about the MS (i.e. in sSFR) at constant $z$ and $M_{\star}$ are associated with variations both in gas to stellar mass ratio ($f_{\rm gs}\propto\Delta_{\mathrm{MS}}^{0.54}$) and in depletion timescale ($t_{\rm dep} \propto \Delta_{\mathrm{MS}}^{-0.43}$), in agreement with observational works by \citet{magdis12}, \citet{sargent14}, \citet{huang14}, \citet{silverman15}, \citet{scoville15}, and G15. Galaxies above the MS ridge have larger gas fractions and smaller depletion times than galaxies at or below the MS ridge. 

The position about the MS has an even stronger dependence on the central gas density ($\rho_{\rm gas,1}\propto\Delta_{\mathrm{MS}}^{0.83}$), i.e., on how the gas is distributed within the galaxies. This may indicate that the central gas density is the key factor involved in determining the MS width, i.e., an internal property has an important role in the confinement of the MS (though it may be stimulated externally, e.g. by a merger). The high central gas densities in more compact SFGs at the top of the MS lead to a decrease in the free-fall time $t_{\rm ff}$, which itself leads to a shorter depletion time ($t_{\mathrm{dep}}\approx t_{\rm ff}/\varepsilon_{\rm ff}$) even for a fixed total gas mass. Already \citet{elbaz11}, \citet{wuyts11}, \citet{lada12}, and \citet{sargent14} put forward the idea that the decrease in $t_{\mathrm{dep}}$ above the MS may be associated with internal parameters such as the central gas density. Furthermore, such compact SFGs (i.e., the blue nuggets) have been observationally detected \citep{barro13, barro14, nelson14_nature, bruce14_size, williams14, williams15}.

\subsection{Turnaround at the Upper and Lower Edge of the MS}

\begin{figure*}
\includegraphics[width=\textwidth]{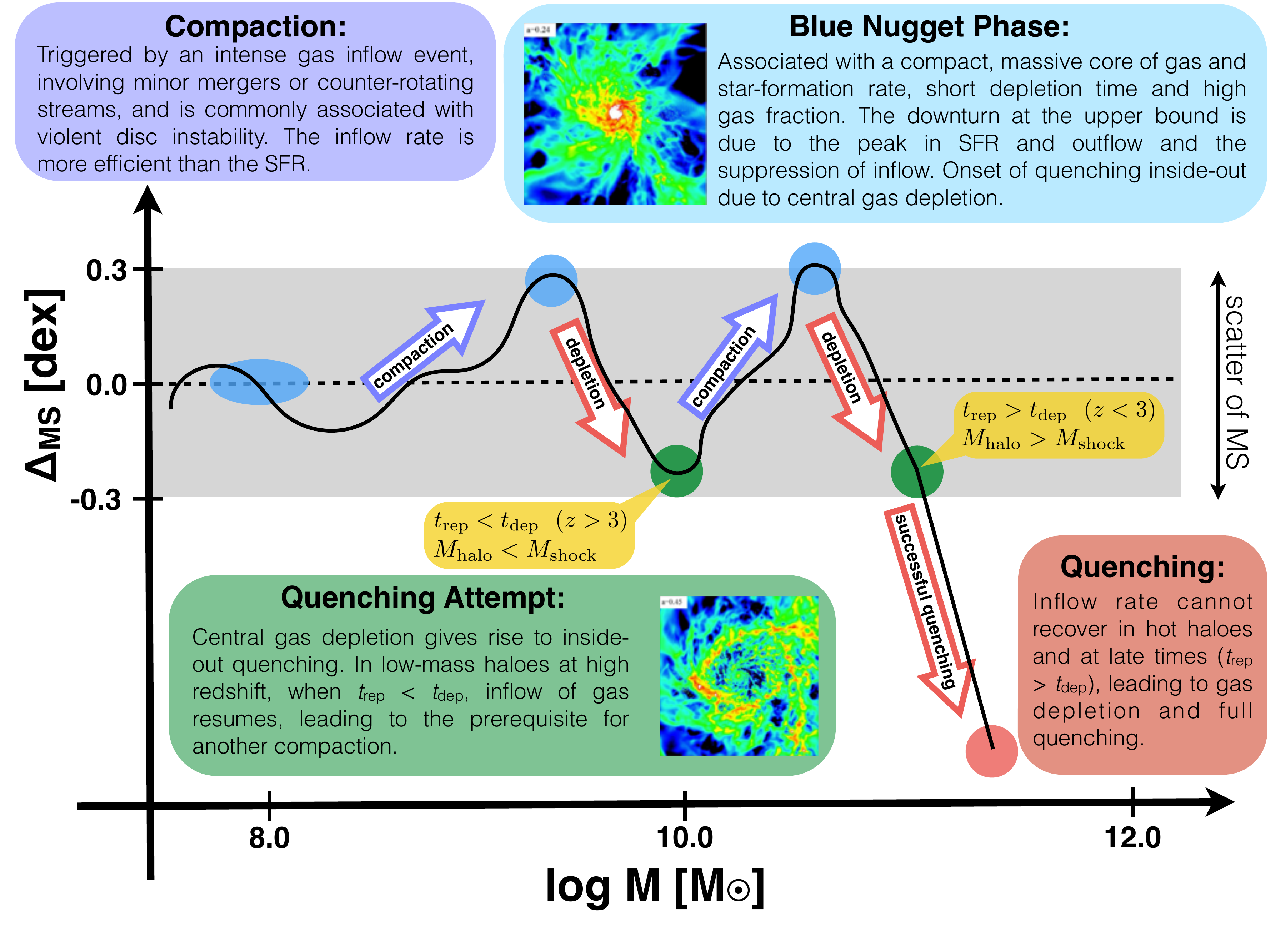} 
\caption{Sketch of the self-regulated evolution along the MS. SFGs are confined to a narrow MS before they quench. During this evolution, the galaxy lives through one or more blue nugget phases during which a minimum in gas depletion time and a maximum in gas fraction are reached. The blue nugget phases are followed by gas depletion inside-out. This quenching attempts fail for low halo masses and at high redshifts since the recovered inflow rate triggers a new episode of compaction and high star formation. At high halo masses (hot halo), the inflow rate cannot recover and the galaxies ceases its star formation activity. }
\label{Fig:MS_Model}
\end{figure*}

As highlighted in Section~\ref{sec:GalaxyProperties}, galaxies oscillate around the MS equilibrium on timescales of $0.4~t_{\rm H}$. Figure~\ref{Fig:MS_Model} sketches the evolution of a typical MS galaxy that eventually quenches its star formation at a late time in the massive end. During the oscillating evolution along the MS, a galaxy reaches minima and maxima in the distance from the MS ($\Delta_{\mathrm{MS}}$). The massive galaxies (as measured at a given time, say $z=2$) typically have one maximum, while less massive galaxies can have more than one. 

As shown before, the climb of a galaxy towards the top of the MS is due to a wet gas compaction, during which the gas inflow rate to the centre is faster than the SFR \citep{dekel14_nugget}. This leads to high SFR in a compact SFG (blue nugget), with high gas fraction and short depletion time. The turnaround at the top of the MS is a natural result of more efficient gas depletion (shorter depletion time) by the high SFR and the associated feedback-driven outflows, as observed by \citet{cicone16}, combined with the suppression of gas inflow into the centre because the gas disc has shrunk and (at least temporarily) disappeared, or became gravitationally stable (morphological quenching, \citealt{martig09, genzel14a, tacchella15_sci}). 

However, not all galaxies fully deplete and quench after the first turnaround at the top of the MS. In galaxies of relatively low stellar mass, the low-mass halo allows rapid replenishment of the disc by fresh gas, with the replenishment time being shorter than the depletion time ($t_{\rm rep}<t_{\rm dep}$). This sets the condition for another wet gas compaction, which can be triggered by mergers, counter-rotating streams or recycled gas, and can be associated with VDIs. The galaxies, now at the lower envelope of the MS after the quenching attempt, turn around towards a new compact blue nugget phase with high SFR at the top of the MS. Once the replenishment becomes inefficient compared to the depletion ($t_{\rm rep}>t_{\rm dep}$), typically at late redshifts and when the halo mass is above the threshold for virial shock heating, the conditions for wet compaction are not recovered. This allows the galaxy to quench inside-out all the way and thus drop below the MS. 

Self-regulation is the emerging feature that explains the small scatter about the MS ridge. Galaxies are not able to shoot above the MS ridge more than a few tenths of dex because the blue nugget phase naturally triggers central depletion, as the gas supply from the disc has been suppressed and the central gas is rapidly consumed by SFR and outflows. On the other hand, at the lower envelope of the MS, the gas inflow quickly recovers, especially at high redshifts and in low halo masses, giving rise to a new compaction episode and an increase in star formation. Furthermore, the depletion and quenching at late cosmic times ($z<2$) and above a critical mass, with no push-back upwards through compaction, explains the downward slope of the MS at the high-mass end \citep[e.g.,][]{elbaz07, whitaker14, schreiber15}.

\subsection{Timescales for the Evolution on the MS} \label{subsec:timescales}

The goal of this section is to estimate the timescales that are key to understand the evolution of galaxies along the MS and the confinement to it, namely the depletion time ($t_{\rm dep}$) and the replenishment time ($t_{\rm rep}$).

\subsubsection{Depletion Time}

As discussed in Section~\ref{subsec:MS_gradients} and shown in Figure~\ref{FigApp:gradient}, in the simulations near the ridge of the MS, we measure an average depletion time of 

\begin{equation}
t_{\mathrm{dep}} = 0.44 \times (1+z)^{-0.39} \times \left( \frac{M_{\star}}{10^{10.5}M_{\odot}} \right)^{-0.19} ~\mathrm{Gyr}.
\label{eq:tdep_estimate}
\end{equation}
\noindent
This value and its dependence on cosmic time are consistent with observational estimates at $z=1-3$ \citep{tacconi10, tacconi13, genzel15}. 

The weak dependence of $t_{\mathrm{dep}}$ on redshift, which is rather surprising at a first glance given the general growth of galactic dynamical times as $(1+z)^{-3/2}$, may be qualitatively understood by the compaction events as follows. As mentioned in Section~\ref{sec:GasContent}, with a constant SFR efficiency $\varepsilon_{\mathrm{ff}}$, the variation in $t_{\mathrm{dep}}$ mostly reflects variations in $t_{\mathrm{ff}}$ \citep{krumholz12a}. In the Toomre regime, valid at high redshift, star formation occurs mostly in giant clumps. In these clumps, for a constant spin parameter for haloes and galaxies (in mass and redshift), one expects a systematic growth of $t_{\mathrm{dep}}$ with cosmic time:

\begin{equation}
t_{\mathrm{dep}} \propto t_{\mathrm{ff}} \propto t_{\mathrm{d}} \propto t_{\mathrm{H}} \propto (1+z)^{-3/2},
\label{eq:tdep_prop_th}
\end{equation} 
\noindent
where $t_d$ is the galaxy dynamical crossing time. This is in contrast with the slow growth of $t_{\mathrm{dep}}$ in the simulations and observations. 

At low redshifts, as argued in \citet{krumholz12a}, there is a transition to the giant molecular cloud regime, where the surface density is constant. Then, the growth of $t_{\mathrm{ff}}$ with time is suppressed, and one expects a similar effect on $t_{\mathrm{dep}}$. However, it turns out that the predicted slowdown in the growth rate of $t_{\rm dep}$ occurs too late and is insufficient for explaining the indicated slow growth of $t_{\rm dep}$.

Another possibility is that the sequence of wet compaction events may provide a clue for the suppressed growth of $t_{\mathrm{dep}}$. Each such event causes a decrease in the dynamical time of the galaxy, and thus a corresponding decrease in $t_{\mathrm{dep}}$, balancing the natural systematic increase in time based on Equation~\ref{eq:tdep_def}. This can be addressed in conjunction with the oscillations in the MS. When passing through the ridge on the way up (increasing $\Delta_{\mathrm{MS}}$) at a later time, at the early stages of a compaction process, the system is still extended, so $t_{\mathrm{dep}} \propto t_{\mathrm{d}}$ is still relatively long, roughly following Equation~\ref{eq:tdep_prop_th}. However, when passing through the MS ridge on the way down (decreasing $\Delta_{\mathrm{MS}}$), during the early stages of the post-compaction quenching process, the system is still gas-rich, with a short $t_{\mathrm{dep}} \propto t_{\mathrm{d}}$. The compaction causes a decrease in $t_{\mathrm{d}}$, which balances the natural increase in time from Equation~\ref{eq:tdep_prop_th}. At high-$z$, there are galaxies moving both up and down the MS. However, at later $z$, more galaxies are at their post-blue-nugget phase, moving down at the MS ridge, allowing the compaction-driven decline of $t_{\mathrm{d}}$ to balance the natural cosmological growth of $t_{\mathrm{d}}$. Since more massive galaxies quench earlier, and to higher densities, we expect their $t_{\mathrm{dep}}$ at the MS ridge to become shorter, and at earlier times, as seen in the simulations. 

Our attempt to address the origins of the time evolution of $t_{\rm dep}$ is only a qualitative preliminary step. What matters for the arguments below concerning the confinement of the MS is the general evolution of $t_{\rm dep}$ with cosmic time and mass. 

\subsubsection{Replenishment Time}

After the central depletion started at the blue nugget phase, it can either bounce back at the lower edge of the MS into a new compaction phase, or it can continue to quench to well below the MS (Figure~\ref{Fig:MS_Model}). The critical times to be compared to the depletion time are the time for replenishment of the gas disc, and the time for the next intense accretion episode, e.g. a merger, that can trigger a new compaction event. As long as the halo is not massive and hot enough to suppress the streaming of cold gas through it, these two timescales can be approximated by the timescale for mass accretion into the galaxy, the inverse of the specific accretion rate given in Equation~\ref{eq:sAR}, namely,

\begin{equation}
t_{\mathrm{rep}} \sim 25 s_{0.04}^{-1}(1+z)^{-2.5}~\mathrm{Gyr} \simeq
  \begin{cases}
   1.60 \mathrm{Gyr} & z=2 \\
   0.78 \mathrm{Gyr} & z=3~, \\
   0.45 \mathrm{Gyr} & z=4
  \end{cases}
\label{eq:trep_estimate}
\end{equation}
\noindent
where $s_{0.04}=s_{\rm h}/0.04\approx1$. 

The replenishment time can become much longer if the halo is more massive than a threshold mass, $M_{\mathrm{vir}} \sim M_{\mathrm{shock}} \sim 10^{11.5}~M_{\odot}$, such that it can support a stable virial shock that keeps the circum-galactic medium at the virial temperature, and when cold streams are suppressed at late redshifts \citep{birnboim03, dekel06}.

\subsubsection{Quenching Attempt versus Full Quenching}\label{subsubsec:attemptVSfull}

\begin{figure}
\includegraphics[width=\linewidth]{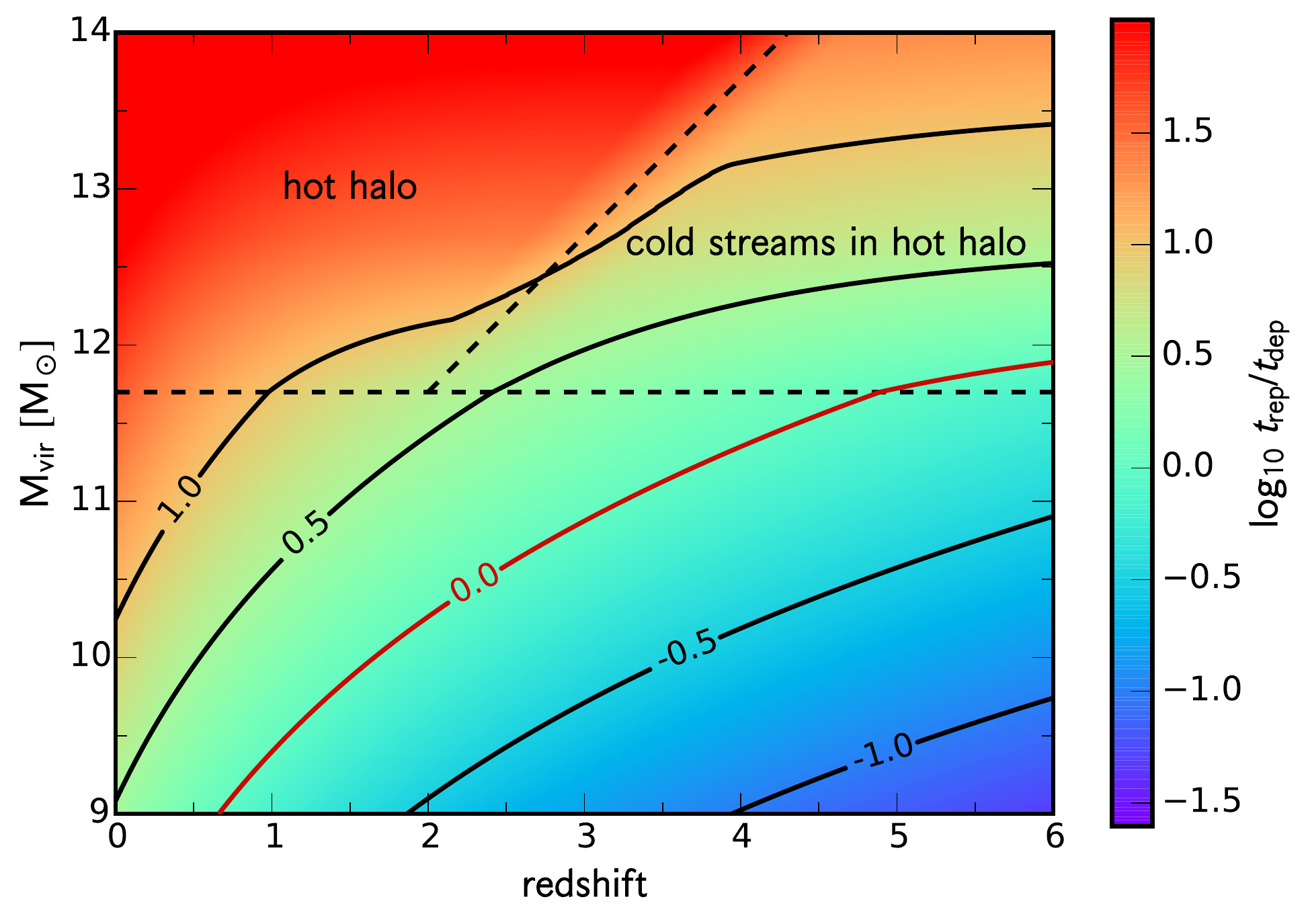} 
\caption{Sketch of expected quenching efficiency in the plane of halo mass $M_{\rm vir}$ and redshift $z$. Colour coding and contours correspond to the expected quenching efficiency, which is given by the ratio of the replenishment time $t_{\rm rep}$ (Equation~\ref{eq:trep_estimate}) and the depletion time $t_{\rm dep}$ (Equation~\ref{eq:tdep_estimate}). The horizontal line at $M_{\rm vir}=10^{11.7}~M_{\odot}$ crudely marks the threshold mass for a stable shock based on spherical infall analysis \citep{dekel06}, which in practice stretches over an order of magnitude in mass. Below this curve the flows are predicted to be predominantly cold and above it a shock-heated medium is expected to extend out to the halo virial radius. The inclined dashed curve is the conjectured upper limit for cold streams in hot haloes, valid at redshifts higher than $z_{\rm crit}\sim2$. }
\label{Fig:Quenching_Analysis}
\end{figure}

As mentioned before, full quenching is achieved when the replenishment time is longer than the depletion time. In Figure~\ref{Fig:Quenching_Analysis}, we show a very rough estimate of the quenching efficiency, defined as the ratio of the estimated replenishment time $t_{\rm rep}$ (Equation~\ref{eq:trep_estimate}) and the estimated depletion time $t_{\rm dep}$ (Equation~\ref{eq:tdep_estimate}), in the plane of the halo mass $M_{\rm vir}$ and redshift $z$. The red line indicates the boundary where $t_{\rm rep}=t_{\rm dep}$. We caution the reader that at $z<1$, Equation~\ref{eq:trep_estimate} gives only a very crude estimate of the replenishment time $t_{\rm rep}$. 

While being above the red line where $t_{\rm rep} \sim t_{\rm dep}$ indicates that the quenching process can possibly proceed, the actual value of $t_{\rm rep}/t_{\rm dep}$ where full, long-term quenching is achieved may be larger, on the order of a few. For one thing, the newly accreted gas might not be available immediately for star formation, which can cause a delay. This delay might be larger for galaxies towards lower redshifts and it could also depend on stellar mass. Thus, in Figure~\ref{Fig:Quenching_Analysis}, long-term quenching can be crudely expected in the red area above $t_{\rm rep}/t_{\rm dep}\sim10$. 

In the regime of $M_{\mathrm{vir}}<M_{\mathrm{shock}}\approx10^{11.7}$, comparing Equation~\ref{eq:tdep_estimate} with Equation~\ref{eq:trep_estimate}, we see that, for a galaxy with a stellar mass of a few times $10^{10}~M_{\odot}$, the condition for full quenching is valid for $z<2.5$. For massive galaxies, the condition is valid earlier ($z<3.0$), whereas for lower-mass galaxies, the condition is valid later ($z<2.0$). This may explain the decisive quenching of massive galaxies at high-$z$, even before halo quenching dominates. In addition, the hot medium in haloes of $M_{\rm vir}>M_{\rm shock}$ at $z>z_{\rm crit}\sim2$ is predicted to host penetrating cold streams, while haloes of a similar mass at $z<z_{\rm crit}$ are expected to be all hot, shutting off most of the gas supply to the inner galaxy. Therefore, once $M_{\mathrm{vir}}>M_{\mathrm{shock}}$, the hot halo can make $t_{\mathrm{rep}}$ much longer, so the condition $t_{\rm rep}>t_{\rm dep}$ is more easily fulfilled even at high redshifts (see also Fig. 7 of \citealt{dekel06}).

\subsection{Further Estimates} \label{subsec:further_estimates}

First, we estimate the timescale for compaction and check whether it is consistent with the Fourier analysis of the MS oscillations presented in Section~\ref{subsec:Oscillation}. Secondly, we calculate the width of the MS based on several simple assumptions. We caution the reader that both the estimate for the compaction timescale and the calculation of the width of the MS are rather crude and should be regarded as consistency checks.

\subsubsection{Compaction Time}

Galaxies do not remain ``sub-'' or ``super-'' MS galaxies, but they rather oscillate about the MS ridge. In the simulations, e.g., Fig. 16 of \citet{zolotov15}, the duration of the compaction phase from the onset of compaction to the blue nugget is on average about

\begin{equation}
t_{\mathrm{com}} \sim (0.3-0.4)t_{\mathrm{H}} \simeq
  \begin{cases}
   1.00 \mathrm{Gyr} & z=2 \\
   0.66 \mathrm{Gyr} & z=3~, \\
   0.50 \mathrm{Gyr} & z=4 
  \end{cases}
\label{eq:tcom}
\end{equation}
\noindent
where $t_{\mathrm{H}}$ is the Hubble time at the blue nugget phase. 

We can very crudely estimate the expected ballpark of $t_{\mathrm{com}}$ in the following two ways. If the compaction phase is driven by a minor merger event, we expect $t_{\mathrm{com}}$ to be comparable to the duration of a minor merger, from the first close passage to coalescence. This is in the ball park of the halo crossing time, $t_{\mathrm{vir}} \sim 0.2~t_{\mathrm{H}}$ (based on spherical collapse), which is not far from what we see in the simulations. 

Alternatively, assuming VDI-driven wet inflow \citep{dekel09, dekel09b, dekel14_nugget}, we can evaluate $t_{\mathrm{com}}$ by the evacuation time of the disc, 

\begin{equation}
t_{\mathrm{com}} \sim \alpha^{-1}t_{\mathrm{mig}} \sim 50t_{\mathrm{d}} \sim 2t_{\mathrm{vir}} \sim 0.4t_{\mathrm{H}},
\end{equation}
\noindent
assuming $\alpha\sim0.2$ for the fraction of cold disc mass in clumps. The timescale $t_{\mathrm{mig}}$ is the migration time of the clumps, and $t_{\mathrm{d}}$ is the disc crossing time, which can be estimated by $t_{\mathrm{d}} \sim \lambda t_{\mathrm{vir}}$ with the spin parameter $\lambda \sim 0.04$. This estimate recovers again the ballpark of $t_{\mathrm{com}}$ from the simulations.

\subsubsection{Oscillation Timescale}

In Section~\ref{subsec:Oscillation}, we measured the timescale for a full oscillation at $z=3-6$ based on a Fourier analysis to be about $0.2-0.5~t_{\rm H}$. We can now estimate this oscillation timescale from the derived timescales for compaction and replenishment. From Equations~\ref{eq:trep_estimate} and \ref{eq:tcom}, we obtain for the oscillation timescale $t_{\rm osc} \approx t_{\rm com} + t_{\rm rep} \approx (0.3 + 0.3)~t_{\rm H} \sim 0.6~t_{\rm H}$, which is roughly consistent with the Fourier analysis.

\subsubsection{The Width of the MS}

In Section~\ref{sec:MainSequence}, we determined $\sigma_{\mathrm{MS}}=0.27$ at $z=3-6$, which is in good agreement with observations. Based on the physical picture above, we can attempt to estimate the expected scatter of the MS from the estimated timescales. 

Consider a galaxy that moves from the upper to the lower edge of the MS over $t_{\mathrm{rep}} \sim t_{\mathrm{dep}}$. Assume that during this phase, the gas mass is about half its peak value at the blue nugget point on average, so the SFR is about half its peak value (by the Kennicutt law). Since there is not much inflow in this phase, the baryon mass is roughly constant. Therefore, if the gas mass is half what it was at the peak, the stellar mass is roughly twice what it was. This means that the sSFR has dropped by more than a factor of 4. During the quenching episode, say between $z=4$ and 3, the sSFR MS ridge has dropped by a factor of $\sim1.6$. This gives $\sigma_{\mathrm{MS}} \ga \log_{10}(\sqrt{4/1.6}) \approx 0.20$ dex, similar to the observed and simulated width of the MS. 

Next, consider a galaxy during its compaction phase, from the lower to the upper edge of the MS. During this phase, on average, the gas density in the centre increases by a factor of order 10, while the total gas mass is changing by a much smaller factor. Based on Equation~\ref{eq:tdep_def}, the SFR increases by a factor of $\sqrt{10} \approx 3.2$. The stellar mass may be doubling its value from the green to the blue nugget point. Therefore, the sSFR increases by a factor of $\sim1.6$. As before, the sSFR MS ridge has dropped during the compaction phase by a factor of $\sim1.6$. This gives $\sigma_{\mathrm{MS}} \ga \log_{10}(\sqrt{1.6\times1.6}) \approx 0.20$ dex, again in the ball park of the observed and simulated width of the MS.

%%%%%%%%%%%%%%%%%%%%%%%%%%%%%%%%%
\section{Discussion}\label{sec:Discussion}

In this section, we discuss the implications of the picture of the MS as discussed above for the quenching of galaxies. Furthermore, we highlight how the outlined picture of the MS can be confirmed in observations. Finally, we highlight some caveats of the analysis presented here and how one can improve in future work. 

\subsection{Implications for Quenching}

There is solid observational evidence that the cessation of star formation in some galaxies, which results in the emergence of quiescent galaxies, correlates with both galaxy mass and environment \citep[e.g.,][]{dressler80, balogh04_bimodality, baldry06, kimm09, peng10_Cont, woo13, knobel13, kovac14}. Furthermore, quenching has been observed to correlate strongly with morphology and galaxy structure in the local universe \citep{kauffmann03, franx08, cibinel13a, fang13, schawinski14, bluck14, woo15} and at high-$z$ \citep{wuyts11, wuyts12, bell12, cheung12, szomoru12, barro13, lang14, tacchella15}. On the other hand, quenched disc galaxies have also been observed \citep{mcgrath08, van-dokkum08, bundy10, salim12, bruce12, carollo14}. 

The physical nature of quenching, and its link to morphology, is a central issue in galaxy evolution. From theory, it has been proposed that galaxies starve out of gas by rapid gas consumption into stars, in combination with the associated outflows driven by stellar feedback \citep[e.g., ][]{dekel86, murray05} or super-massive black hole \citep[e.g.,][]{di-matteo05, croton06, ciotti07, cattaneo09}, and/or by a slowdown of gas supply into the galaxy \citep[e.g.,][]{rees77, dekel06, hearin13, feldmann15_quench}. Another possibility includes morphological quenching, which argues that the growth of a central mass concentration, i.e., a massive bulge, stabilizes a gas disc against fragmentation \citep{martig09}. 

Using observations alone, it is difficult to constrain the physical nature of quenching because a correlation between quenching and a certain galaxy property does not necessarily imply a direct causal relation or the direction of such a causality \citep[e.g.,][]{carollo14}. For example the observed correlation between central stellar density and quenching could either arise because high stellar density causes quenching, or because another property that is associated with high stellar density causes quenching, or because quenching leads to high stellar density. The other property may be, for example, central gas density, AGN feedback, or halo mass. As an example, Lilly \& Carollo (2016, in preparation) show that a central surface density threshold for quenching could in principle be the result of a hierarchical accretion of mass in galaxies, in which galaxies that quench earlier are denser.

\citet{tacchella15_sci} mapped out the $M_{\star}$ and SFR distribution on scales of 1 kpc in $z\sim2$ SFGs. They found that the $\sim10^{11}~M_{\odot}$ galaxies quench inside-out, where the star formation in the centre ceases within $\la200$ Myr after $z\sim2$, whereas the outskirts still form stars for $1-3$ Gyr. In our simulations, we find a similar inside-out quenching signature where the depletion starts from the centre in the post-blue-nugget phase (see also \citealt{tacchella15_profile}, for a detailed analysis of the evolution for the $M_{\star}$ and SFR profiles in the simulations). As shown above, our simulated galaxies are able to deplete and quench rapidly at $z\sim1-3$, forming galaxies that resemble today's typical $\sim10^{11}~M_{\odot}$ galaxies, which show features of a gas-rich, dissipative formation process \citep[e.g.,][]{bender88, carollo93, faber97, cappellari07}. 

From the analysis of the simulations presented here, we can now make a step forward in understanding the physical nature of quenching. Our simulations show that the interplay between the gas and the stellar components together with the dark-matter halo is able to explain quenching of galaxies. Our analysis highlighted the key role of halo mass, which directly translates to the gas replenishment time of the galaxy's disc (see Section~\ref{subsubsec:attemptVSfull}). We have seen that galaxies depleted their centres first, i.e., quenching progresses inside-out. These observational signatures have been found also in observations of $z\sim2$ galaxies. Furthermore, the observation that the central stellar mass density (bulge mass) is a good indicator for quiescence \citep{kauffmann03, franx08, cheung12, fang13, bluck14, lang14, woo15, barro15_structure} is in good agreement with the thoughts presented here. If a compaction with a nuclear ``starburst'' precedes quenching, one expects a high central stellar mass density to have built up. Similarly, due to the key role of halo mass in our picture, and the correlation of stellar mass with halo mass, we would also expect that total stellar mass is a good measure for quiescence \citep[e.g.,][]{peng10_Cont, carollo13a}.

\subsection{Observational Consequences}

As demonstrated above, our simulations reveal gradients of gas fraction and depletion time across the MS that are consistent with the observed gradients by G15, and explain the phenomena by the evolution through compaction, depletion, and quenching events. Our additional key prediction is a strong gradient of the core gas density, e.g. $\rho_{\rm gas,1}$. Future observations that will resolve the gas distribution within individual galaxies should be able to confirm this. Furthermore, star formation is expected to occur centrally concentrated at the top of the MS, whereas it is predicted to be ring-like distributed in massive systems at the lower envelope of the MS proceeding the blue-nugget phase before quenching. Indications for this have already been seen in \citet{genzel14a} and \citet{tacchella15_sci}.

\subsection{Caveats}

We have mentioned several limitations of the simulations used here in Section~\ref{subsec:limitations}. For example, our simulations do not include AGN feedback and feedback associated with cosmic rays and magnetic field. Nevertheless, \citet{ceverino14_radfeed} have shown that our simulations match basic observations, such as SFRs, gas fractions, and stellar-to-halo mass fractions, at least at least within observational uncertainties. If the SFR at very high redshifts is still overestimated, the dramatic events in the evolution of galaxies that concern us here may occur somewhat earlier than in the real universe. However, it seems that in the current simulations these evens, the compaction and the subsequent onset of quenching, do occur at cosmological times that are consistent with observations. 

The number of simulated galaxies analysed in this work is limited to 26, outputted in $\sim900$ snapshots from $z=6-1$. Clearly, a larger number of simulated galaxies would have been better for studying the average properties of galaxies as a function of time and mass, and the scatter about them, but the number of galaxies is limited being computationally expensive. However, the key aspect of this work is that we resolve the gas physics of high-density star-forming regions at $\sim25$ pc scales in a cosmological context, which allows us to capture the evolution of galaxies along and about the MS. This is key to understand the evolution of galaxies on the MS.

%%%%%%%%%%%%%%%%%%%%%%%%%%%%%%%%%
\section{Conclusion}\label{sec:Conclusion}

Our zoom-in cosmological simulations of relatively massive galaxies at $z>1$ reveal that the SFGs are confined to a universal MS as they evolve through gas compaction, depletion, and replenishment. They reproduce the observed properties of the MS and allow us to present a simple understanding for the evolution and the scatter of the MS. 

In the simulations, in the redshift range $z=6-1$, the MS ridge of sSFR declines in time, sSFR$_{\rm MS} \propto (1+z)^{2.5}$, with only a weak mass dependence. This follows closely the predicted decline of the specific halo-mass accretion rate in the Einstien-deSitter phase of the expanding universe. We find that galaxies oscillate about this MS ridge on timescales of $\sim0.2-0.5~t_{\rm H}$, which corresponds to $0.2-0.9$ Gyr at $z=6-3$. The SFGs are confined to a narrow MS of width $\pm0.27~\mathrm{dex}$. The simulated evolution with redshift, the MS width, and the bending of the MS at high masses are consistent with observations. 

The simulations reveal that the high-SFR galaxies at the upper envelope of the MS tend to be compact blue nuggets with high gas fractions and short depletion times. The lower-SFR galaxies at the lower envelope of the MS typically get there after partial central gas depletion that is naturally triggered at the blue-nugget phase, and they have lower gas fractions and longer depletion times. The measured gradients of gas fraction and depletion times across the MS are in agreement with observations. The steepest gradient across the MS is predicted for the central gas density, $\rho_{\rm gas,1}$, which indicates that the main mechanism responsible for the width of the MS involves an internal property of the galaxy, though it may be determined by an external process. On the other hand, we do not find any significant gradients across the MS for stellar structure properties, namely central stellar mass surface density, size, and S\'{e}rsic index. 

Propagation up towards the upper envelope of the MS (blue nugget phase) is associated with gas compaction of the gas disc, triggered by a minor merger or counter-rotating streams through violent disc instability, where the inflow rate is higher than the SFR. The blue nugget phase marks the onset of gas depletion, during which the central gas is exhausted due to SFR and outflow, while the inflow from the disc that has shrunk is suppressed. In the post-compaction phase the galaxies show the signature of inside-out quenching, where the SFR first reduces in the centre. After reaching the bottom of the MS, an upturn can occur if the extended disc has been replenished by fresh gas and a new compaction event is triggered prior to total gas depletion, namely, if $t_{\rm rep} < t_{\rm dep}$. The self-regulated nature of these mechanisms can explain the confinement of SFGs into a MS narrower than $\pm 0.3$ dex.

Full quenching with a departure from the MS occurs in massive haloes or at low redshifts, where the disc replenishment is slow compared to the depletion time, which we find to be $0.2-1.0~\mathrm{Gyr}$, declining with mass and increasing with cosmic time, consistent with observations. The requirement for full quenching, $t_{\rm dep} < t_{\rm rep}$, tends to be fulfilled at redshifts $z\leq3$, and possibly at higher redshifts for more massive galaxies. A halo above the critical mass for virial shock heating, $\sim10^{11.5}M_{\odot}$, helps suppressing the replenishment and causing long-term quenching.

\section*{Acknowledgements}

We acknowledge stimulating discussions with Guillermo Barro, Sandy Faber, Reinhard Genzel, Mark Krumholz, Simon Lilly, Benny Trakhtenbrot, Joanna Woo, and Adi Zolotov. We thank the referee for constructive and useful comments, which helped to improve the manuscript. ST thanks AD and his group for the hospitality during his visit at HUJI. Development and most of the analysis have been performed in the astro cluster at HU. The simulations were performed at the National Energy Research Scientific Computing Center (NERSC), Lawrence Berkeley National Laboratory, and at NASA Advanced Supercomputing (NAS) at NASA Ames Reserach Center. This work was supported by ISF grant 24/12, by GIF grant G-1052-104.7/2009, by the I-CORE Program of the PBC, ISF grant 1829/12, by MINECO grant AYA2012-32295, by CANDELS grant HST-GO-12060.12-A, and by NSF grants AST-1010033 and AST-1405962. We acknowledge support by the Swiss National Science Foundation.

\appendix
\section{Stellar-to-halo mass relation}
\label{App:stellarhalo}

\begin{figure}
\includegraphics[width=\linewidth]{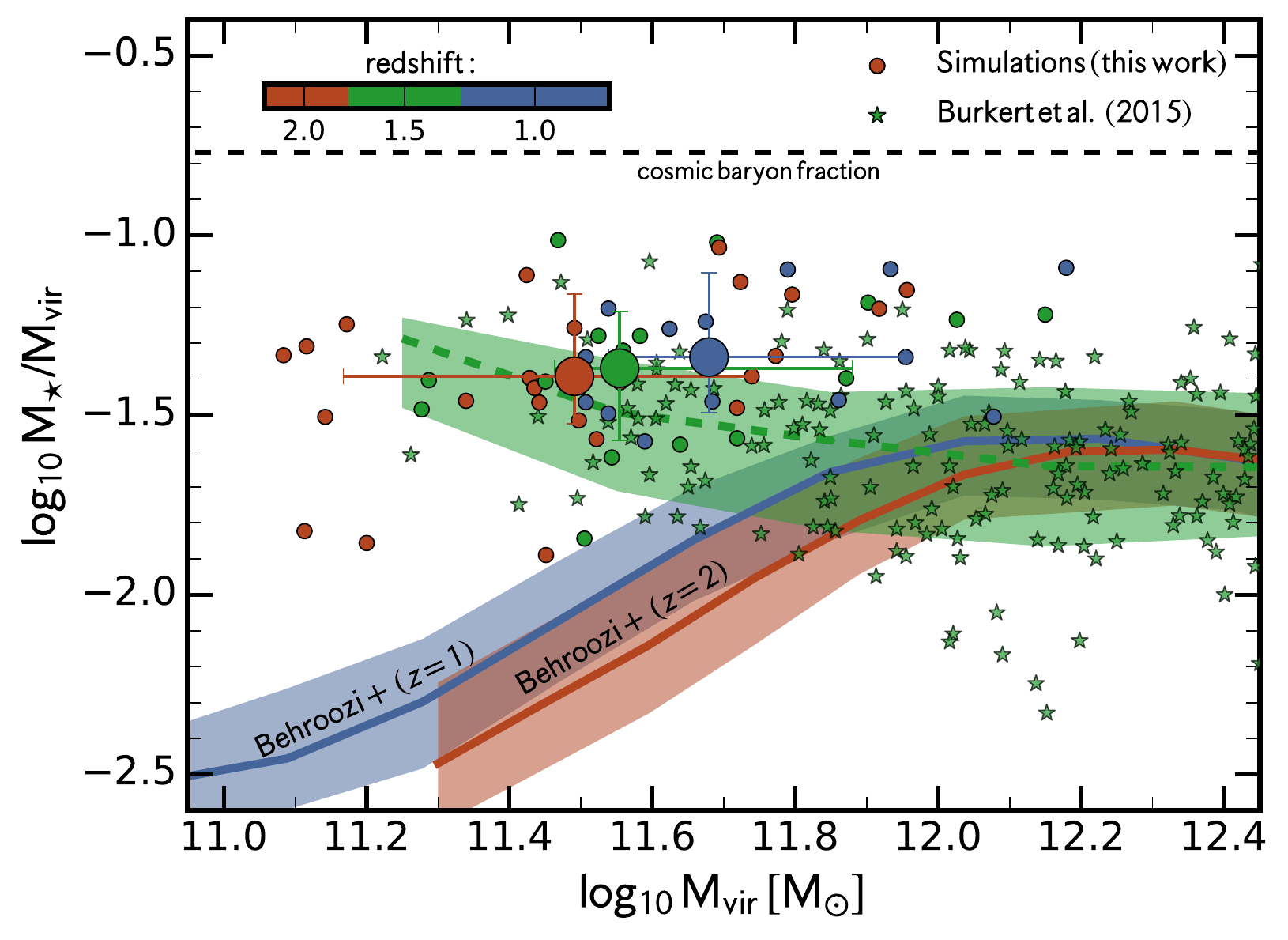} 
\caption{Ratio of stellar mass to halo mass as a function of halo mass. The red, green and blue point show our simulations at $z=2$, 1.5 and 1.0, respectively. The large symbols show the median (at its $1~\sigma$ scatter) at these three redshifts. The horizontal black dashed line shows the cosmic baryon fraction of the Universe. The solid lines show the \citep{behroozi13b} relations. Our simulations lie a factor of $2-6$ above the \citet{behroozi13b} relation at $z\sim1-2$, but are in agreement with recent observations of SFGs at $0.5\la z \la 2.6$ \citep{burkert15}, which are shown as stars and their median as dashed line.}
\label{FigApp:stellarhalo}
\end{figure}

One of the key quantity for simulations to match is the stellar mass made within a given dark-matter halo. In this section, we compare the stellar-to-halo mass ($M_{\star}-M_{\rm vir}$) relation of the simulations with results from abundance matching \citep{conroy09, moster10, moster13, behroozi10, behroozi13b} and observed kinematics \citep{burkert15}. We have already presented a similar comparison in our companion paper \citep{tacchella15_profile}.

Specifically, we compare our simulations with data from \citet{behroozi13b} and \citet{burkert15}. \citet{behroozi13b} derive the $M_{\star}-M_{\rm vir}$ relation by first assuming a general form of this relation, and then obtains the parameters by simultaneously performing abundance matching of the mass functions at different redshifts and comparing the consequent information on star formation with a variety of observational data, such as the evolution of the MS and the global SFR density of the Universe. \citet{burkert15} derive dark-matter halo masses from recent observations of the H$\alpha$ kinematics of $z\sim0.8-2.6$ galaxies, i.e., these estimates are totally independent of abundance matching. To derive the halo mass, they assume implicitly that the specific angular momentum of the baryons on the scale of the dark halo is the same as that of the dark matter component. Hence, their results depend on the angular momentum distribution on the halo scale (of both baryons and dark matter), as well as on any re-distribution of angular momentum between different baryonic components (e.g., inner and outer disk, outflow, bulge).

Figure~\ref{FigApp:stellarhalo} shows the ratio of stellar mass to halo mass as a function of halo mass for our simulations and the estimates from observational data \citep{behroozi13b, burkert15}. At $z=1$, the simulated galaxies have a median halo mass of $\log_{10}~M_{\rm vir}=11.7\pm0.3$ and a stellar to halo mass ratio of $\log_{10}~M_{\star}/M_{\rm vir}=-1.3\pm0.2$. \citet{burkert15} found $\log_{10}~M_{\star}/M_{\rm vir}=-1.5\pm0.3$ at $\log_{10}~M_{\rm vir}=11.7$, which is consistent with our simulations, while \citet{behroozi13b} found $\log_{10}~M_{\star}/M_{\rm vir}=-1.8\pm0.2$ via abundance matching, which is a factor of 3 lower than our simulated estimate. At $z=2$, we find only little evolution in the halo-to-stellar mass ratio: our simulated galaxies have a median halo mass of $\log_{10}~M_{\rm vir}=11.5\pm0.3$ and a stellar to halo mass ratio of $\log_{10}~M_{\star}/M_{\rm vir}=-1.4\pm0.2$. Via abundance matching, \citet{behroozi13b} found $\log_{10}~M_{\star}/M_{\rm vir}=-2.3\pm0.2$ at $\log_{10}~M_{\rm vir}=11.5$, but \citet{burkert15} found $\log_{10}~M_{\star}/M_{\rm vir}=-1.6\pm0.3$, which is again consistent with our simulations.

We conclude that our simulations produce stellar to halo mass ratios that are in the ballpark of the values estimated from observations, and within the observational uncertainties. Therefore, it is sensible and adequate to use our simulations with the adopted feedback prescription, while bearing in mind the factor-of-two uncertainties.

\section{Analysis of thin Timesteps}
\label{App:thin}

\begin{figure*}
\includegraphics[width=\textwidth]{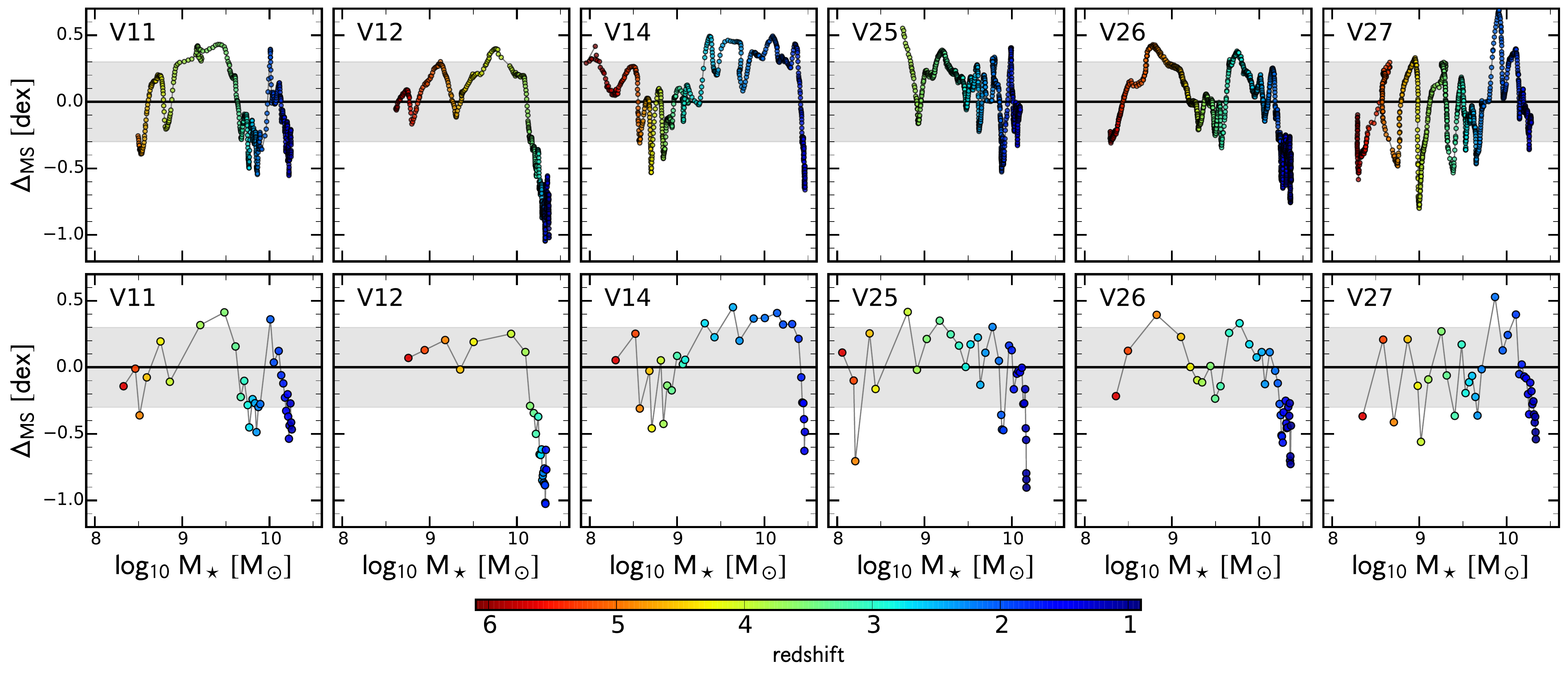} 
\caption{Evolutionary tracks about the universal MS comparing short and standard timesteps between snapshots. We compare for six galaxies the higher resolution time-steps (top panels, snapshots separated by $\Delta a\approx0.0006$) with the standard resolution time-steps (bottom panels, $\Delta a\approx0.01$). We find that the standard resolution tracks the main features of the evolution along the MS rather well.}
\label{FigApp:thin}
\end{figure*}

For six galaxies (11, 12, 14, 25, 26, and 27), there are thinner time-steps available for the analysis. Our standard temporal resolution is $\Delta a=0.01$), which corresponds to $\sim100$ Myr between each snapshot. This is roughly half the orbital time at the disc edge, and therefore it should be short enough to trace galaxy internal processes. 

We nevertheless compare our obtained results with this standard temporal resolution with the high resolution. The thinner time-steps are separated by $\Delta a=0.0005-0.0007$, they have a $\sim20$-times higher resolution. In Figure~\ref{FigApp:thin}, we plot the evolution along the MS (distance from the MS versus total stellar mass) for the higher (upper panels) and standard (bottom panels) resolution time-steps. We find that the standard resolution tracks the higher resolution very well: all the main features over long ($\sim1$ Gyr) as well as short ($\sim100$ Myr) are there. Only the very short term fluctuations ($<100$ Myr) are underestimated, leading to a deficit of power of short-term changes (see also Section~\ref{subsec:Oscillation}).

\section{Gradients across the MS}
\label{App:gradient}

One of the main goals of this paper is to determine the galaxy properties across the MS. As discussed in Section~\ref{sec:GalaxyProperties}, if we want to determine these MS gradients, we have to take into account that galaxy properties may also evolve with cosmic time. We therefore have to correct for the systemic time evolution of galaxies near the MS ridge. 

\subsection{Gas-related gradients}

\begin{figure*}
\includegraphics[width=\textwidth]{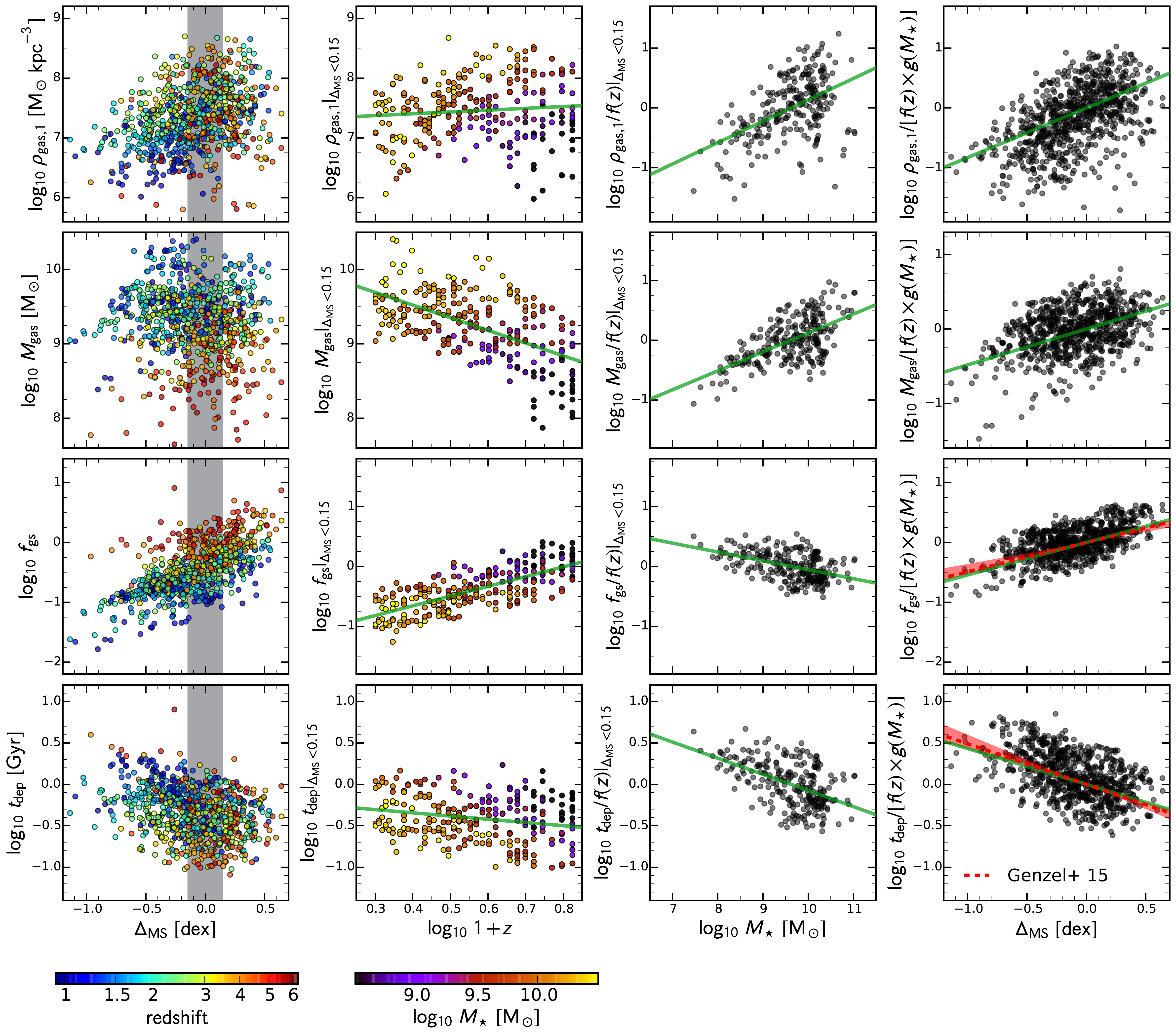} 
\caption{Galaxy properties across the MS. Shown from top to bottom are the gas mass within 1 kpc, total gas mass, gas mass to stellar mass ratio, and depletion time. In the left panels, we plot these quantities as a function of the distance from the MS $\Delta_{\mathrm{MS}}$. The redshift dependence is colour coded. The middle-left panels show the intrinsic redshift dependencies, colour-coded by stellar mass. The green line indicates the best-fitting redshift dependence $f(z)$. The middle-right panels show the redshift-corrected quantities as a function of stellar mass. The green line indicates the best-fitting mass dependence $g(z)$. The right panels display the quantities corrected for redshift and mass (same as Figure~\ref{Fig:Gradients_MS}). The green line indicates the best-fitting, while the red dashed line shows the best-fit of the observations by G15.}
\label{FigApp:gradient}
\end{figure*}

Figure~\ref{FigApp:gradient} is a more extended version of Figure~\ref{Fig:Gradients_MS}. We again investigate the following four key quantities are investigated from top to bottom: the central gas density within 1 kpc ($\rho_{\rm gas,1}$), the total gas mass ($M_{\rm gas}$), gas to stellar mass ratio ($f_{\rm gs}$), and depletion time ($t_{\mathrm{dep}}$). 

The left column shows these (uncorrected) quantities as a function of the distance from the MS ridge ($\Delta_{\mathrm{MS}}$) for all galaxies at $z=1-6$. The colour coding of the points corresponds to redshift. We confirm the trends found in Section~\ref{subsec:MS_gradients}: $\rho_{\rm gas,1}$ increases systematically by about one order of magnitude from below to above the main sequence, i.e., as $\Delta_{\mathrm{MS}}$ varies from $-0.5$ to $+0.5$ dex. The trend for the total $M_{\rm gas}$ is weaker, partly contaminated by high-redshift galaxies with low gas mass near the MS ridge. $f_{\rm gs}$ shows not only a correlation with $\Delta_{\mathrm{MS}}$, but also a clear redshift evolution which has been highlighted also in Figure~\ref{Fig:GasMass}. The depletion time $t_{\mathrm{dep}}$ shows a negative correlation with $\Delta_{\mathrm{MS}}$, with several outliers with high $t_{\mathrm{dep}}$ at $z>3$. Those points indicate the quenching attempts which were not successful, due to the high redshift and a too low halo mass (see Section~\ref{subsubsec:attemptVSfull} below). 

The best fits for the $z$-dependence ($f(z)=\xi+\zeta\times\log_{10}(1+z)$) are shown in the middle-left panels of Figure~\ref{FigApp:gradient} with the colour-coding corresponding to stellar mass $M_{\star}$. We find a steep redshift evolution for $M_{\rm gas}$ and $f_{\rm gs}$ with $\zeta=-1.71\pm0.15$ and $+1.63\pm0.10$, respectively. For $\rho_{\rm gas,1}$ and $t_{\mathrm{dep}}$, the $z$-evolution is much shallower with $\zeta=+0.30\pm0.21$ and $-0.39\pm0.11$, respectively (Table~\ref{tab:best_fit_gradients}). 

In a second step, we correct for the stellar mass dependence, which is shown in the middle-right panels of Figure~\ref{Fig:Gradients_MS}. All four quantities depend on $M_{\star}$, largely due to the strong correlation between gas fraction $f_{\rm gas}$ and $M_{\star}$ (see Section~\ref{sec:GasContent} and Figure~\ref{Fig:GasMass}). We fit the $M_{\star}$-dependence of the $z$-corrected quantities with the following relation: $g(M_{\star})=\eta+\gamma\times(\log_{10}(M_{\star})-10.5)$. We find $\gamma=+0.36\pm0.04$, $+0.32\pm0.04$, $-0.15\pm0.02$, and $-0.19\pm0.02$ for $\rho_{\rm gas,1}$, $M_{\rm gas}$, $M_{\rm gas}/M_{\odot}$, and $t_{\mathrm{dep}}$, respectively (Table~\ref{tab:best_fit_gradients}). 

The right panels of Figure~\ref{FigApp:gradient} show the key quantities ($\rho_{\rm gas,1}$, $M_{\rm gas}$, $f_{\rm gs}$, and $t_{\mathrm{dep}}$) corrected for their systematic $z$-evolution $f(z)$ and $M_{\star}$-dependence $g(M_{\star})$ (same as Figure~\ref{Fig:Gradients_MS}). By construction, we find for all key quantities a tighter correlation than before the correction. Fitting these corrected quantities with $Q=\alpha+\delta\times\Delta_{\mathrm{MS}}$, we find $\delta=+0.83\pm0.04$, $\delta=+0.48\pm0.03$, $\delta=+0.54\pm0.02$, and $\delta=-0.43\pm0.02$ for $\rho_{\rm gas,1}$, $M_{\rm gas}$, $f_{\rm gs}$, and $t_{\mathrm{dep}}$, respectively (Table~\ref{tab:best_fit_gradients}). Interestingly, the gradient of $\rho_{\rm gas,1}$ across the MS is the steepest. This demonstrates that galaxies above the MS ridge have a significantly higher central gas density than galaxies below it: going from $\Delta_{\rm MS}=0.5$ dex below the MS ridge to 0.5 dex above it, we find that the $\rho_{\rm gas,1}$ increases by nearly an order of magnitude.

\subsection{Stellar-structure gradients}

\begin{figure*}
\includegraphics[width=\textwidth]{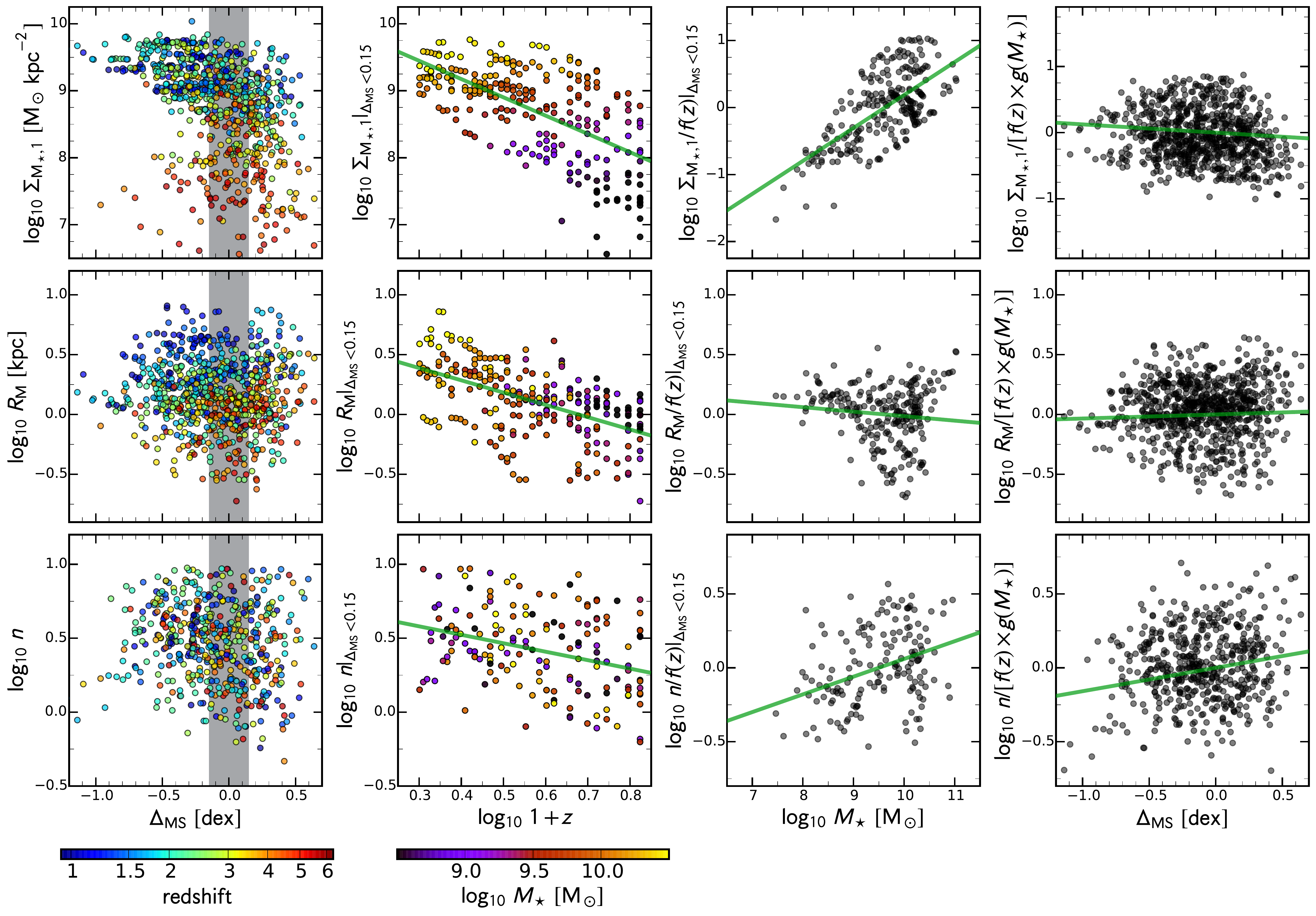} 
\caption{Same as Figure~\ref{FigApp:gradient}, but for the stellar structure quantities central stellar mass density within 1 kpc ($\Sigma_{M_{\star},1}$, half-mass radius ($R_{\rm e}$), and S\'{e}rsic index ($n$).}
\label{FigApp:gradient_structure}
\end{figure*}

Following the same approach as before for the gas-related gradients across the MS, we focus here on the stellar-structure gradients. Figure~\ref{FigApp:gradient_structure} is a more extended version of Figure~\ref{Fig:Gradients_MS_structure}, showing the central stellar mass density within 1 kpc ($\Sigma_{M_{\star},1}$, the half-mass radius ($R_{\rm e}$), and the S\'{e}rsic index ($n$). From the left to the right, we show the uncorrected quantities, the $z$-dependence, the $M_{\star}$-dependence, and the corrected quantities. The best-fitting values are given in Table~\ref{tab:best_fit_gradients}. The main difference to the gas-related gradients is that there is no correlation between the stellar-structure quantities and the distance to the MS, and the inferred gradients are much shallower.

\section{Inflow Rate, Outflow Rate and Star-Formation Rate along the MS}
\label{App:driver}

\begin{figure*}
\includegraphics[width=\textwidth]{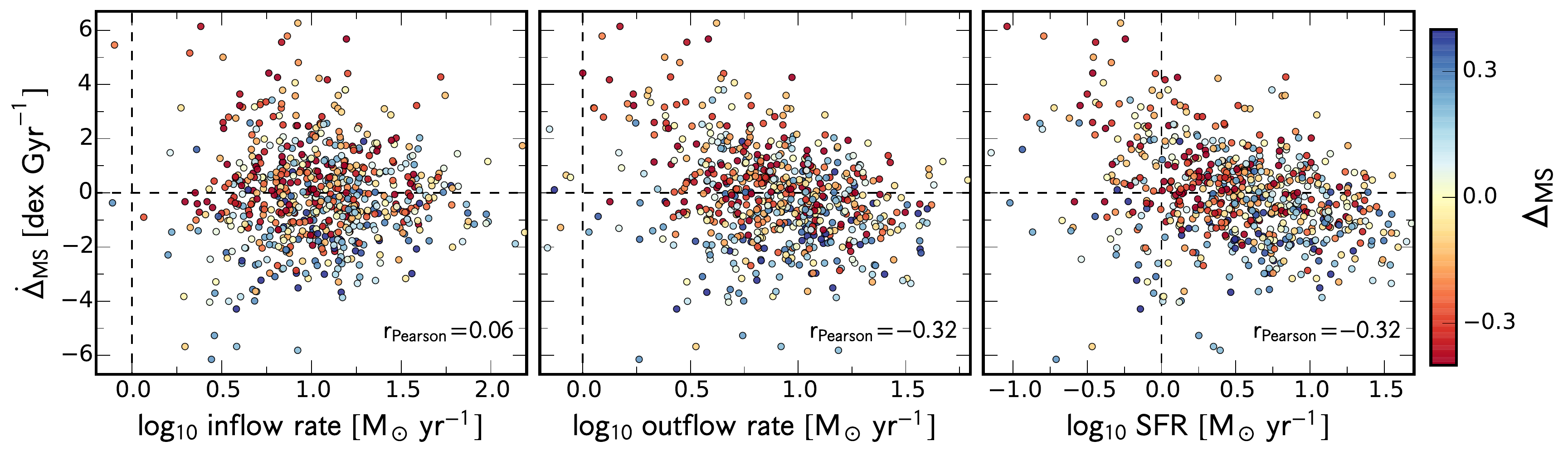} 
\caption{The origin of the evolution across the MS. The three panels show $\dot{\Delta}_{\rm MS}$ as a function of the inflow rate, outflow rate, and SFR from left to right, i.e., we have split the balance term $\log_{10}~B_{5\mathrm{kpc}}$ of Figure~\ref{Fig:Balance}. The colour coding corresponds to the position on the MS, $\Delta_{\rm MS}$. We find that the outflow rate and SFR are moderately correlated with $\dot{\Delta}_{\rm MS}$, whereas the inflow rate is not correlated with $\dot{\Delta}_{\rm MS}$. }
\label{FigApp:BalanceMS}
\end{figure*}

In Section~\ref{subsec:Driver} we discuss the driver of the MS gradients, focusing on the balance between the gas inflow rate (input term) and SFR plus gas outflow rate (drainage term). Figure~\ref{Fig:Balance} shows the relation between the rate of change of distance from the MS, $\dot{\Delta}_{\rm MS}$, and the balance between the input term and drainage terms of gas in the central 5 kpc, namely $\log_{10}~B_{5\mathrm{kpc}}=\log_{10}[\mathrm{inflow~rate}/(\mathrm{SFR}+\mathrm{outflow~rate})]$. Here, we split the balance term $\log_{10}~B_{5\mathrm{kpc}}$ into its constituents, namely inflow rate, outflow rate, and SFR. 

Figure~\ref{FigApp:BalanceMS} shows the change of the distance from the MS, $\dot{\Delta}_{\rm MS}$, as a function of inflow rate, outflow rate, and SFR, each measured within the central 5 kpc. We find that the individual rate are correlated to a lesser degree than the combined balance term $B_{5\mathrm{kpc}}$. The outflow rate and SFR are moderately correlated with $\dot{\Delta}_{\rm MS}$ ($r_{\rm Pearson}=-0.32$), while the inflow rate is not correlated with $\dot{\Delta}_{\rm MS}$.

\bsp

\label{lastpage}

\end{document}